\documentclass[conference]{IEEEtran}
\IEEEoverridecommandlockouts
\usepackage{amssymb}
\usepackage{hyperref}
\usepackage{cite}
\usepackage{graphicx}
\usepackage{array}
\usepackage{multirow}
\usepackage{amsmath}
\usepackage{stfloats}
\usepackage{graphicx}
\usepackage{subfigure}
\usepackage{tabularx}
\usepackage{epsfig,epsf}
\usepackage{setspace}
\usepackage{bm}
\usepackage{textcomp}
\usepackage{algorithmic}
\usepackage{algorithm}
\usepackage{subfigure}
\usepackage{caption}
\usepackage{graphicx}
\usepackage{setspace}
\usepackage{url}
\usepackage{verbatim}
\usepackage{graphicx}
\usepackage{cite}
\usepackage{amssymb}
\usepackage{color}
\usepackage{xpatch}
\definecolor{shadecolor}{RGB}{0,0,255}
\definecolor{blue}{RGB}{0,0,255}
\newtheorem{theorem}{Theorem}

\newtheorem{corollary}{Corollary}
\newtheorem{remark}{Remark}

\makeatletter

\ExplSyntaxOn
\cs_new:Npn \bibColoredItems #1#2
{
	\clist_map_inline:nn {#2} { \cs_new:cpn {bib@colored@##1} {#1} } 
}
\ExplSyntaxOff

\newcommand\bib@setcolor[1]{%
	\ifcsname bib@colored@#1\endcsname
	\expanded{\noexpand\color{\csname bib@colored@#1\endcsname}}%
	\else
	\normalcolor
	\fi
}

\makeatother

\begin{document}
	
\title{From Active to Battery-Free: Rydberg Atomic Quantum Receivers for Self-Sustained SWIPT-MIMO Networks}

\author{Qihao Peng, 
          Qu Luo,~\IEEEmembership{Member,~IEEE},
           Zheng Chu,~\IEEEmembership{Member,~IEEE},
           Neng Ye,~\IEEEmembership{Member,~IEEE},\\
        Hong Ren,~\IEEEmembership{Member,~IEEE}, 
        Cunhua Pan,~\IEEEmembership{Senior Member,~IEEE},
        Lixia Xiao,~\IEEEmembership{Member,~IEEE},\\
        Pei Xiao,~\IEEEmembership{Senior Member,~IEEE}.
  %      Chau Yuen,~\IEEEmembership{Fellow,~IEEE}.
   
		\thanks{Q. Peng, Q. Luo, and P. Xiao are affiliated with 5G and 6G Innovation Centre, Institute for Communication Systems (ICS) of the University of Surrey, Guildford, GU2 7XH, UK. (e-mail: \{q.peng, q.u.luo,p.xiao\}@surrey.ac.uk). } \\
        \thanks{Z. Chu is with the Department of Electrical and Electronic Engineering, University of Nottingham Ningbo, China. (e-mail: andrew.chuzheng7@gmail.com).}
       \thanks{N. Ye is with the School of Cyberspace Science and Technology, Beijing Institute of Technology, Beijing 100081, China. (e-mail: ianye@bit.edu.cn).}
        \thanks{L. Xiao is with the Research Center of 6G Mobile Communications, the School of Cyber Science and Engineering, Huazhong University of Science and Technology, Wuhan 430074. (e-mail: lixiaxiao@hust.edu.cn).}
          \thanks{H. Ren and C. Pan are with the National Mobile Communications Research Laboratory, Southeast University, Nanjing, China. (e-mail:  \{hren,cpan\}@seu.edu.cn).} }
       % \thanks{T. Gong and C. Yuen are with the School of Electrical and Electronics Engineering, Nanyang Technological University, Singapore 639798 (e-mail: trgTerry1113@gmail.com, chau.yuen@ntu.edu.sg). }}

            %} 
	
	\maketitle

\begin{abstract}
In this paper, we proposed a hybrid simultaneous wireless information and power transfer (SWIPT)–enabled multiple-input multiple-output (MIMO) architecture, where the base station (BS) uses a conventional RF transmitter for downlink transmission and a Rydberg atomic quantum receiver (RAQR) for receiving uplink signal from Internet of Things (IoT) devices. To fully exploit this integration, we jointly design the transmission scheme and the power-splitting strategy to maximize the sum rate, which leads to a non-convex problem. To address this challenge, we first derive closed-form lower bounds on the uplink achievable rates for maximum ratio combining (MRC) and zero-forcing (ZF), as well as on the downlink rate and harvested energy for maximum ratio transmission (MRT) and ZF precoding. Building upon these bounds, we propose an iterative algorithm relying on the best monomial approximation and geometric programming (GP) to solve the non-convex problem. Finally, simulations validate the tightness of our derived lower bounds and demonstrate the superiority of the proposed algorithm over benchmark schemes. Importantly, by integrating RAQR with SWIPT-enabled MIMO, the BS can reliably detect weak uplink signals from IoT devices powered only by harvested energy, enabling battery-free communication.
\end{abstract}	
	

\section{Introduction}
A central pillar of the 5G and envisioned 6G paradigm is ubiquitous connectivity, driving an unprecedented scale of interconnection and functional diversity \cite{andrews2014will,wang2023road}. A fundamental challenge in realizing this vision lies in powering the massive Internet of Things (IoT) devices, as traditional approaches relying on battery replacement or wired charging are impractical for devices deployed in inaccessible locations or large-scale deployments. Simultaneous wireless information and power transfer (SWIPT) provides a promising solution for this challenge by leveraging electromagnetic waves to deliver both information and energy simultaneously \cite{perera2017simultaneous,pan2017performance}, enabling massive connectivity for 6G. 

Owing to the appealing features, SWIPT has been widely studied in emerging technologies, including multiple-input multiple-output (MIMO) \cite{tang2017energy,tang2017joint,zhang2015large}, cell-free massive MIMO \cite{femenias2021swipt,hua2025cell}, reconfigurable intelligent surface (RIS) \cite{pan2020intelligent,Yang2024,li2021joint}, unmanned aerial vehicle (UAV) \cite{feng2020uav,Zhang2024}, and advanced multiple access techniques \cite{xu2017joint,tang2019energy}. Leveraging advanced MIMO beamforming, a base station (BS) can deliver RF energy toward intended targets, which improves both wireless power transfer efficiency and information throughput \cite{tang2017energy}. To further exploit the benefits of MIMO, the antenna selection and spatial switching schemes have been designed to enhance energy efficiency \cite{tang2017joint}. In addition, the secrecy rate of MIMO-SWIPT systems has been investigated under statistical channel state information (CSI) to ensure confidentiality for information-decoding users \cite{zhang2015large}. Considering the severe interference from multiple cells, cell-free massive MIMO has been shown to provide uniform service quality to multiple users without cell boundary \cite{ngo2017cell,peng2022resource}. With this architecture, the power control coefficients were obtained to maximize the minimum weighted rate in the SWIPT-enhanced cell-free massive MIMO system \cite{femenias2021swipt}. The harvested energy was further maximized by optimizing the access point (AP) selection and power control \cite{hua2025cell}. Unlike the active APs, deploying the power-free RIS can enhance both harvested energy and achievable data rates by tuning a large number of passive elements \cite{pan2020intelligent,Yang2024}. Then, the joint beamforming and power splitting was developed for RIS-enabled SWIPT networks \cite{li2021joint}. In addition to RIS, UAVs can simultaneously expand coverage areas to establish reliable connections. For example, \cite{feng2020uav} studied the benefits of UAV-enabled SWIPT systems. Zhang \emph{el al.} \cite{Zhang2024} further unlocked the potential of UAV in satellite-UAV-terrestrial systems.
Regarding non-orthogonal multiple access (NOMA), a cooperative protocol was proposed in \cite{xu2017joint}, where the user with good channel conditions acts as a relay to enhance the performance of user with poor CSI. Similarly, the joint power allocation and time switching control were optimized to improve energy efficiency of NOMA-SWIPT system \cite{tang2019energy}. 
Despite significant advances in SWIPT, relying only on harvested energy to maintain reliable communication for IoT devices is still challenging, as the available energy inherently decays with propagation distance.

By leveraging quantum phenomena to achieve unprecedented precision in measuring physical quantities, quantum sensing provides a promising solution for detecting weak RF signals \cite{hanzo2025quantum,gisin2007quantum,degen2017quantum}. Particularly, Rydberg atoms exhibit exaggerated atomic properties such as large electric dipole moments and strong polarizability \cite{sedlacek2012microwave}, enabling them to interact with external electromagnetic fields across a broad frequency range from direct current (DC) to terahertz (THz) bands. Furthermore, by exploiting quantum effects such as electromagnetically induced transparency (EIT) \cite{finkelstein2023practical} and Autler–Townes splitting (ATS) \cite{hao2018transition}, an incident weak RF signal can be detected by a photodetector. These remarkable capabilities effectively bridge the quantum domain and classical wireless communication. By enabling RF-to-optical conversion with quantum-limited sensitivity, the Rydberg atomic quantum receivers (RAQRs) exhibit ultra-high sensitivity, broadband tunability, narrowband selectivity, and broad coverage.

Building upon these appealing characteristics, extensive researchers have explored the benefits of RAQR in communication \cite{gong2024rydberg,gong2025rydberguplink,zhu2025general,cui2025towards} and sensing \cite{chen2025polarization}. Specifically, an equivalent baseband model for single-input single-output (SISO) RAQR was presented in \cite{gong2024rydberg}, quantifying the considerable performance gain over conventional RF chains. This model was subsequently extended to a MIMO architecture \cite{gong2025rydberguplink}, demonstrating the superiority of RQAR over conventional MIMO systems. To capture time-varying signals, a dynamic-response model of RAQRs was proposed in \cite{zhu2025general}. In contrast to the equivalent complex channel models in \cite{gong2024rydberg,gong2025rydberguplink,zhu2025general}, the authors of \cite{cui2025towards} investigated a magnitude-only model and developed the corresponding signal detection algorithms. Based on this magnitude-only model, the advantages of RAQRs for sensing have also been explored in \cite{chen2025polarization}. Even though prior studies have demonstrated the potential of RAQRs, two fundamental issues remain unresolved. One is whether RAQRs can detect ultra-weak uplink signals enabled by the harvested energy, thereby enabling battery-free communication. The other is how the presence of RAQRs reshapes the transmission design in practice, including pilot and payload power allocation, power splitting, and transmission strategies.

To the best of our knowledge, this paper is the first of its kind to integrate RAQRs into SWIPT-enabled MIMO systems and to evaluate its feasibility for battery-free communication at the IoT device side. Specifically, the main contributions of this work are summarized as follows:
\begin{itemize}
	\item We investigate a SWIPT–MIMO framework that combines a conventional RF transmitter with a RAQR. Under this framework, the BS transmits information and energy in the downlink, while the uplink pilots and data, powered by the harvested energy, are collected by the RAQR. Leveraging the high sensitivity of the RAQR, extremely weak RF signals can be reliably detected, which significantly improves channel estimation accuracy and data detection performance. 
	
	\item  Within this architecture, we first evaluate channel estimation performance with RAQRs and derive closed-form lower bounds for the achievable rates and the harvested energy under linear precoding and detection. These analytical results reveal the fundamental tradeoffs among downlink energy allocation, uplink pilot power, and uplink payload power. Furthermore, Monte-Carlo simulations validate the accuracy of the derived bounds, proving that the lower bounds can provide tractable expressions for system design.
	\item Building on these bounds, we formulate a joint uplink–downlink optimization that maximizes the sum rate subject to SWIPT and feasibility constraints. To solve the resulting non-convex problem, the alternative optimization is employed to divide the original problem into two sub-problems, namely, joint transmission design and block optimization. Then, the best monomial approximation is adopted to transform the joint transmission design into a geometric programming (GP) problem, while block optimization is addressed via mixed-integer programming. Furthermore, a rigorous convergence analysis is provided to prove that our proposed algorithm can converge to a locally optimal solution.

	\item We conduct extensive simulations to validate that the  proposed algorithm consistently outperforms the benchmark schemes in terms of sum rate, especially in the low-power regime, thereby enabling battery-free device operation. These results indicate that integrating RAQR with SWIPT-enabled MIMO provides a promising solution for battery-free IoT devices without sacrificing spectral efficiency.
\end{itemize}

%Based on this model, both the mean square error (MSE) and normalized mean square error (NMSE) of RAQRs are investigated, and the lower bounds on the achievable rates for maximum-ratio combining (MRC) and zero-forcing (ZF) schemes are derived, respectively. Monte-Carlo simulations validate the theoretical analysis and provide key insights, showing that the performance gains of RAQ-MIMO are primarily driven by the quantum-enhanced sensitivity, especially for the low-power regimes and limited pilot resources. These findings not only deepen the theoretical understanding of quantum-assisted MIMO but also offer practical guidance for pilot design and receiver configuration in future quantum-enhanced wireless communication systems.

\emph{Organization and Notations}: The system model is presented in Section II. The channel estimation and performance analysis are provided in Section III. In Section IV and Section V, the transmission strategies for various precoding/detection schemes are presented. Finally, the simulation and our conclusions are given in Section VI and Section VII, respectively. The notations are given in the following: the superscripts \((\cdot)^T\), \((\cdot)^H\), and \((\cdot)^*\) denote transpose, Hermitian (conjugate transpose), and complex conjugation, respectively. $\mathrm{diag}(\cdot)$ denotes the diagonal matrix formed from a vector; \(\mathbf{I}_M\) means the \(M \times M\) identity.
$\mathbb{R}$, $\mathbb{C}$, and $\mathbb{N}$ represent the real, complex, and natural fields.  $\mathcal{CN}(\boldsymbol{\mu},\boldsymbol{\Sigma})$ is a circularly symmetric complex Gaussian distribution with mean $\boldsymbol{\mu}$ and covariance $\boldsymbol{\Sigma}$.
$\mathbb{E}\{\cdot\}$ denotes expectation.

\section{System Model of RAQ-MIMO}
In this section, we briefly introduce the principles of RAQ-based receiver, and then present the SWIPT-MIMO with the aid of RAQRs.

\begin{figure*}
	\centering
	\includegraphics[width=7in]{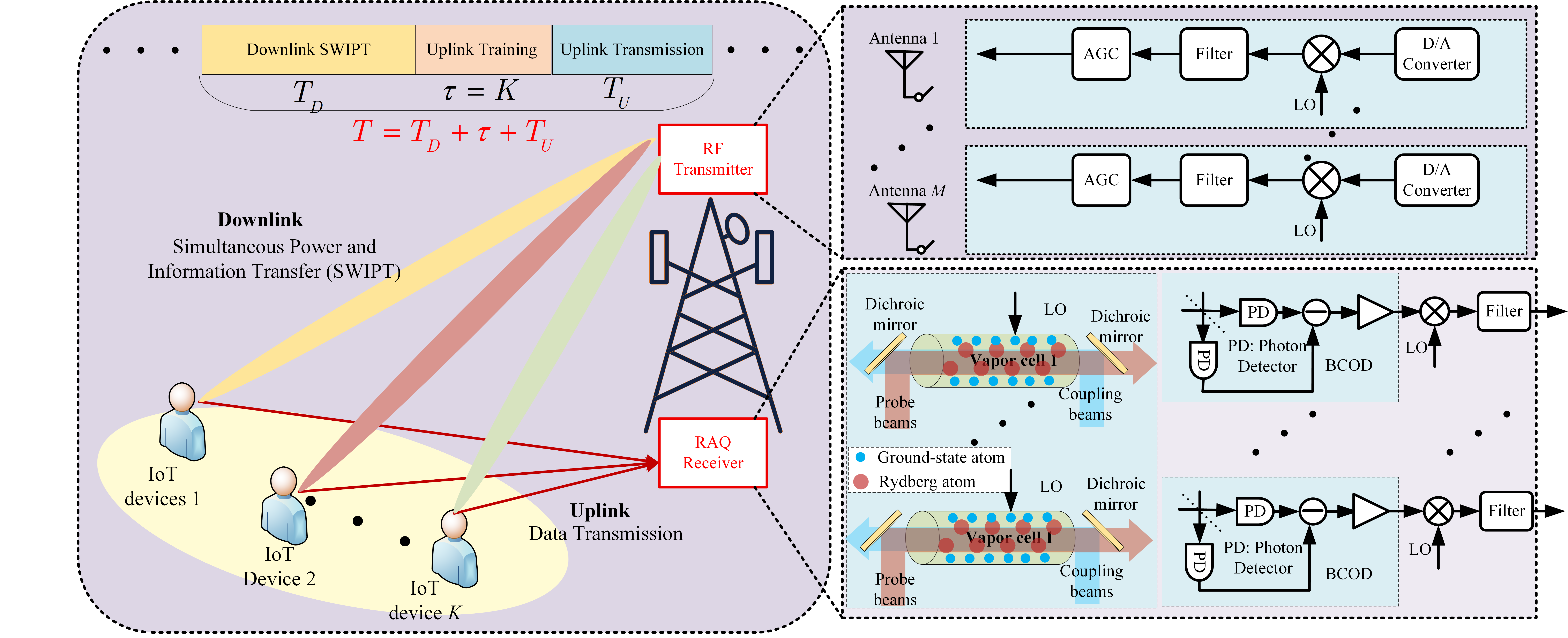}
	\caption{SWIPT-enabled MIMO system with a hybrid architecture, where RF transmitter with \(M\) antennas sends the power and information to \(K\) IoT devices and RAQR with \(M\) atomic cells receive the signal from \(K\) IoT devices.}
	\label{system}
\end{figure*}
\vspace{-0.2cm}

\subsection{System model of RAQ-MIMO}
As illustrated in Fig.~\ref{system}, the BS consists of one RF transmitter equipped with \(M\) antennas and a RAQR with \(M\) vapor cells. Leveraging downlink simultaneous wireless information and power transfer, the BS enables \(K\) IoT devices to harvest energy for uplink training and data transmission. Furthermore, we assume time division duplexing transmission, and thus the uplink and downlink channels are reciprocal.
 
Then, we briefly introduce the system model of RAQ-MIMO, which can be expressed as \cite{gong2024rydberg,gong2025rydberguplink}
\begin{equation}
    \label{systemmodel}
    \small
    \mathbf{y} = \sqrt{\rho}\Phi\mathbf{D}\mathbf{H}\mathbf{s} + \mathbf{n},
\end{equation}
where \(\mathbf{H} = [\mathbf{h}_1,\cdots,\mathbf{h}_K] \in \mathbb{C}^{M \times K}\) is the complex channel, \( \mathbf{n} \sim \mathcal{CN}(\mathbf{0},\sigma^2\mathbf{I}_M)\) means the complex additive white Gaussian noise (AWGN), \(\Phi \triangleq \frac{e^{-j(\theta_l - \varphi(\Omega_l))}}{2} +\frac{e^{-j(\theta_l + \varphi(\Omega_l))}}{2} \) denotes the phase shifts related to the first vapor cell's phase \(\theta_{l} \triangleq \theta_{1,l} \) and phase shift caused by signal superposition \(\varphi(\Omega_l)\) (\cite[Eq.~(12)]{gong2025rydberguplink}), \(\rho\triangleq \rho_m, \forall m\) is the effective gain, and \(\mathbf{D} = \text{diag}\{1,e^{-j\frac{2\pi d\sin\vartheta}{\lambda}},\cdots, e^{-j\frac{2\pi (M-1)d\sin\vartheta}{\lambda}}\}\) is the phase shift matrix. \(\vartheta\), \(d\), and \(\lambda\) are the angle of arrival of the local oscillator (LO) signal, the distance between each vapor cell, and the wavelength, respectively. Furthermore, the channel between the IoT device  \(k\) and the BS \(\mathbf{h}_k \sim \mathcal{CN}(\mathbf{0},\beta_k\mathbf{I}_M)\) follows the complex Gaussian distribution of zero mean and variance of \(\beta_k\). 

\subsection{Channel Estimation}
The transmission mechanism consists of three stages, including uplink training of duration \(\tau = K\), uplink data transmission of duration \(T_{U}\), and downlink SWIPT of duration \(T_{D}\). Furthermore, the coherence time of each frame is fixed, which is given by \(T = \tau + T_{U} + T_{D} \).

Based on the system model in (\ref{systemmodel}), the received pilot signal at the RAQR can be expressed as
\begin{equation}
	\label{pilot}
	\small
	\mathbf{Y} = \sqrt{\rho}\Phi\mathbf{D}\mathbf{H}\mathbf{P}\mathbf{Q}^H + \mathbf{N},
\end{equation}
where \(\mathbf{N} \in \mathbb{C}^{M \times \tau}\) is the background noise, \(\mathbf{P} = \text{diag}\{\sqrt{\tau p^p_1},\cdots,\sqrt{\tau p^p_K}\}\) stands for the pilot power, and \(\mathbf{Q} = [\mathbf{q}_1,\cdots,\mathbf{q}_K]  \in \mathbb{C}^{\tau \times K}\) denotes the pilot sequence with \(\mathbf{q}^H_k\mathbf{q}_{k'} = 0\) when \(k \neq k'\) and \(\mathbf{q}^H_k\mathbf{q}_{k'} = 1\) when \(k = k'\). By multiplying \(\mathbf{q}_k\), we obtain
\begin{equation}
	\label{rxpilot}
	\small
	\mathbf{y}_k =  \sqrt{\rho \tau p^p_k}\Phi\mathbf{D}\mathbf{h}_k + \mathbf{N}\mathbf{q}_k.
\end{equation}
Then, the estimated channel based on the minimum mean square error (MMSE) estimator is given by
\begin{equation}
	\label{estimatedchannel}
	\small
	\begin{split}
		\mathbf{\hat h}_k &= \sqrt{\rho \tau p^p_k}\beta_k\Phi^*\mathbf{D}^H(\rho \tau p^p_k\beta_k|\Phi|^2\mathbf{I}_M+\sigma^2\mathbf{I}_M)^{-1}\mathbf{y}_k \\
		& = \frac{\rho\beta_k \tau p^p_k|\Phi|^2}{\rho\beta_k \tau p^p_k|\Phi|^2+\sigma^2}\mathbf{h}_k + \frac{\sqrt{\rho \tau p^p_k}\beta_k\Phi^*}{\rho\beta_k \tau p^p_k|\Phi|^2 + \sigma^2}\mathbf{D}^H\mathbf{N}\mathbf{q}_k.
	\end{split}
\end{equation}
The MSE matrix and normalized mean square error (NMSE) between \(\mathbf{\hat h}_k\) and \(\mathbf{h}_k\) can be respectively expressed as
\begin{equation}
	\label{MSE}
	\small
	\mathbf{MSE}_k = \frac{\beta_k\sigma^2}{\rho \tau p^p_k\beta_k|\Phi|^2 + \sigma^2}\mathbf{I}_M \triangleq e_k \mathbf{I}_M,
\end{equation}
and 
\begin{equation}
	\label{NMSE}
	\small
	\text{NMSE}_k = \frac{\sigma^2}{\rho \tau p^p_k\beta_k|\Phi|^2 + \sigma^2}.
\end{equation}

By denoting equivalent signal-to-noise ratio (SNR) as \(\gamma_k \triangleq \frac{\rho p^p_k \beta_k |\Phi|^2}{\sigma^2} \), we have \(\text{MSE}_k = \frac{M \beta_k}{1+\tau \gamma_k}\) and \(\text{NMSE}_k = \frac{1}{1+\tau \gamma_k}\), which confirms that RAQRs and MIMO systems have similar channel characteristics, and that the accuracy of channel estimation depends on both channel quality and pilot length. However, considering the unique characteristics of RAQRs, we show the superiority of RAQR over convention RF MIMO system.

\begin{remark}
    \emph{Compared to conventional MIMO, the pilot power can be reduced by a factor of \(10\lg\Big(\frac{\rho|\Phi|^2/\sigma^2}{\rho_{\text{RF}} /\sigma^2_{\text{RF}}}\Big)\), where \(\rho_{\text{RF}}\) and \(\sigma^2_{\text{RF}}\) are the effective gain and noise background of RF MIMO, respectively}.
\end{remark}

Owing to the limited harvested energy, all IoT devices generally work in the \emph{low-energy} regime. Additionally, after accounting for data transmission and circuit consumption, only a small fraction of the harvested energy remains available for pilot transmission. Consequently, both pilot duration and pilot power are constrained, which degrades CSI quality for conventional RF MIMO. By contrast, RAQRs remain advantageous in exactly this regime, since their higher effective pilot SNR enables reliable channel estimation with short and weak pilots, thereby preserving data throughput with limited energy budget.
\subsection{Uplink Transmission}
By assuming that the BS only knows the statistical CSI,  the BS equipped with a RAQR can receive the signal from the \(k\)-th device, which can be expressed as
\begin{equation}
	\label{rxsignal}
	\small
	\begin{split}
		%r_k &= \sqrt{\rho p_k}\Phi\mathbf{c}^{H}_k\mathbf{D}\mathbf{h}_k + \sum\limits_{k' \neq k}^K\sqrt{\rho p_{k'}}\Phi\mathbf{c}^{H}_k\mathbf{D}\mathbf{h}_{k'} + \mathbf{c}^{H}_k\mathbf{w} \\
		y^{U}_k& = \underbrace{ \mathbb{E}\{\sqrt{\rho p^d_k}\Phi\mathbf{c}^{H}_k\mathbf{D}\mathbf{h}_k\} s^{U}_k}_{\text{Ds}_k} \\
		&+ \underbrace{ \sqrt{\rho p^d_k}\Phi\mathbf{c}^{H}_k\mathbf{D}\mathbf{h}_ks^{U}_k -\mathbb{E}\{\sqrt{\rho p^d_k}\Phi\mathbf{c}^{H}_k\mathbf{D}\mathbf{h}_k\}s^{U}_k}_{\text{Ls}_k}  \\
		& + \underbrace{\sum\limits_{k' \neq k}^K\sqrt{\rho p^d_{k'}}\Phi\mathbf{c}^{H}_k\mathbf{D}\mathbf{h}_{k'}s^{U}_{k'}}_{\text{UI}_{k,k'}}+\underbrace{\mathbf{c}^{H}_k\mathbf{n}^{U}_k }_{\text{N}_k},
	\end{split}
\end{equation}
where \(s^{U}_k\) means the information from the \(k\)-th device, \(\mathbf{c}_k \in \mathbb{C}^{M\times 1}\) denotes the \(k\)-th column of the procoding matrix \(\mathbf{C} \in \mathbb{C}^{M \times K}\), and \(\mathbf{n}^{U}_k \sim \mathcal{CN}(\mathbf{0},\sigma^2\mathbf{I}_M)\) represents the Gaussian noise with zero mean and variance of \(\sigma^2\). \(p^d_k\), \(\text{Ds}_k\), \(\text{Ls}_k\), \(\text{UI}_{k,k'}\), and \(\text{N}_k\) represent the \(k\)-th device's transmission power, effective signal, leaked signal, inter-device interference, and noise, respectively. Furthermore, the transmitted data are assumed to be mutually independent. As a result, the achievable data rate \(R_k\) can be expressed as
\begin{equation}
	\label{achRate}
	R^{U}_k = \frac{T_U}{T}
	\log_2\Bigg(1+\frac{|\text{Ds}_k|^2}{|\text{Ls}_k|^2 + \sum\limits_{k' \neq k}^K|\text{UI}_{k,k'}|^2 + |\text{N}_k|^2}\Bigg).
\end{equation}

\subsection{Downlink SWIPT}
The BS simultaneously transmits the information and energy to all devices during the downlink SWIPT. Specifically, the received signal of the device \(k\) can be written as
\begin{equation}
    y^{D}_k = \mathbf{h}^H_k\sum\limits_{k' = 1}^K\sqrt{\rho_{\text{RF}}p^s_{k'}} \mathbf{w}_{k'}s^D_{k'} + n^{D}_k,
\end{equation}
where \(\rho_{\text{RF}}\) is the effective gain of RF receiver, \(p^s_{k'}\) denotes the power dedicated for SWIPT transmission, \(\mathbf{w}_{k'} \in \mathbb{C}^{M \times 1}\) denotes the precoding vector intended for the \(k'\)-th device, \( n^{D}_k \sim (0,\sigma^2_{\text{RF},k})\) means the noise background, and \(s^D_{k'} \) represents transmitted data from the BS to the $k$-th device, which are assumed to be mutually independent, i.e., \(\mathbb{E}\{(s^D_k)^*s^D_{k'}\} = 1\) only when \(k = k'\). Based on the received signal of \(y^{D}_k\), the device \(k\) adopts a power-splitter to achieve SWIPT that can separate the received signal into two parts. One part is utilized for decoding information, while the remaining part is used for energy harvesting (EH) which is utilized for uplink training and transmission. By denoting \(\alpha_k\) as the power-splitting coefficient that defines the amount of information by device \(k\), the energy harvest and decoded information of device \(k\) can be respectively shown as
\begin{equation}
    \small 
    \begin{split}
        y^{\text{EH}}_k = \sqrt{1 - \alpha_k}\Big [\sum\limits_{k' = 1}^K\mathbf{h}^H_k\sqrt\rho_{\text{RF}}{p^s_{k'}} \mathbf{w}_{k'}s^D_{k'} + n^{D}_k\Big],
    \end{split}
\end{equation}
and 
\begin{equation}
    \small 
    \begin{split}
        y^{\text{DI}}_k = \sqrt{\alpha_k}\Big [\sum\limits_{k' = 1}^K\mathbf{h}^H_k\sqrt{\rho_{\text{RF}}p^s_{k'}} \mathbf{w}_{k'}s^D_{k'} + n^{D}_k\Big] + n^D_{k,s},
    \end{split}
\end{equation}
where \(n^D_{k,s} \sim \mathcal{CN}(0,\sigma^2_{k,s})\) is the noise of base-band signal processing and independent with \(n^{D}_k\). 

Through the practical EH circuits, it is challenging to harvest the noise power, and thus the harvested energy can be written as \cite{khodamoradi2022energy}
\begin{equation}
	E_k = \frac{T_D}{T}\eta_{\text{EH}}(1-\alpha_k)\mathbb{E}\Big\{|\mathbf{ h}^H_k\sum\limits_{k' = 1}^K\sqrt{\rho_{\text{RF}}p^s_{k'}}\mathbf{w}_{k'}|^2\Big\},
\end{equation}
where \(\eta_{\text{EH}} \in (0,1]\) is the energy conversion efficiency. The dowlink rate is expressed as
\begin{equation}
	R^{D}_k =  \frac{T_D}{T}B \log_2(1+\text{SINR}^{D}_k),
\end{equation}
where \(\text{SINR}^{D}_k\) can be expressed as (\ref{SINR_down}), which is given at the bottom of the next page.
\begin{figure*}[b]
	\hrule
	\begin{equation}\label{SINR_down}
		\text{SINR}^{D}_k = \frac{\alpha_k\rho_{\text{RF}}p^s_k|\mathbb{E}\{\mathbf{h}^H_k\mathbf{w}_k\}|^2}{\alpha_k\Big( \sum\limits_{k' =1}^K\rho_{\text{RF}}p^s_{k'}\mathbb{E}\{|\mathbf{h}^H_k\mathbf{w}_{k'}|^2\}-\rho_{\text{RF}}p^s_{k}|\mathbb{E}\{\mathbf{h}^H_k\mathbf{w}_k\}|^2 + |n^D_k|^2\Big)+|n^d_{k,s}|^2 }.
	\end{equation}
\end{figure*}

\subsection{Problem Formulation}
By jointly designing uplink and downlink transmission blocks and transmission power, our goal is to maximize both uplink and downlink throughput while meeting devices' requirements. Mathematically, the problem can be formulated as
\begin{subequations}
	\setlength\abovedisplayskip{5pt}
	\setlength\belowdisplayskip{5pt}
	\label{Optimization}
	\begin{align}
		\small
		\mathop {\max }\limits_{\substack{\left\{ \alpha_k \right\},\left\{ {p_k^d} \right\},\left\{p^p_{k}\right\},\\
				\left\{ p^s_{k} \right\}, T_U, T_D}} \quad & \sum\limits_{k=1}^K (R^U_k + R^D_k)\notag\\
		{\rm{s}}{\rm{.t}}{\rm{.}}\;\;\;\; & {R}_k^{U} \ge R_k^{{\rm{req}},U},\forall k, \label{Optimization_A}\\
		& {R}_k^{D} \ge R_k^{{\rm{req}},D},\forall k, \label{Optimization_B} \\
			& \sum\limits_{k=1}^K p^s_k \le P^{s,\max}, \label{Optimization_C} \\
						& K p^p_k + T_U p^d_k \le T E_k, \forall k, \label{Optimization_D} \\
			& T_U + T_D \le T - K, \label{Optimization_F} \\
			&0 \le \alpha_k \le 1, \forall k, \label{Optimization_E} \\
			& T_U,T_D \in \mathbb{N}, \label{Optimization_G}
	\end{align}
\end{subequations}
where \(R_k^{{\rm{req}},U}\) and \(R_k^{{\rm{req}},D}\) are the required data rate for uplink and downlink, respectively. \(P^{s,\max}\) denotes the maximum transmission power of the BS. Constraints \eqref{Optimization_A} and \eqref{Optimization_B} ensure that each device's required rate is satisfied. Constraint \eqref{Optimization_C} enforces the BS transmit-power limit. Constraint \eqref{Optimization_D} means that the uplink energy consumption not exceed the harvested energy \(E_k\). Finally, constraints \eqref{Optimization_F} and \eqref{Optimization_G} ensure that the total block duration should be integers and does not exceed \(T\) and constraint (\ref{Optimization_E}) restricts the power splitting coefficient to the interval [0,1].

Problem~\eqref{Optimization} is challenging to solve as the sum-rate maximization is an NP-hard problem. In addition, the coupling between uplink payload power and pilot power, both of which are constrained by the harvested energy, further increases the intractability of this problem.

\section{Energy Harvesting and Achievable Data Rate}
In this section, we assume that the BS adopts the linear precoding/detection for downlink SWIPT and uplink detection to make Problem (\ref{Optimization}) more tractable. Then, the closed-form expressions for harvested energy and achievable donwlink/uplink rate are derived.

\subsection{Closed-Expression of Uplink Achievable Rate}
By utilizing the maximum rate combination (MRC) and zero-forcing (ZF) detection, we have
\begin{equation}
	 \mathbf{C} = \begin{cases}
	 	\Phi\mathbf{D \hat H}, & \text{for } \text{MRC},\\
	 	\Phi\mathbf{D}\mathbf{\hat H} [(\Phi\mathbf{D}\mathbf{\hat H})^H\Phi\mathbf{D}\mathbf{\hat H}]^{-1}, & \text{for } \text{ZF}.
	 \end{cases}
\end{equation}
Then, we derive the lower bound of uplink data rate with the following theorems.

\begin{theorem}
	\label{MRC_SINR_T}
	The ergodic achievable rate for the $k$-th device using the MRC decoder can be lower bounded by
	\begin{equation}
		\setlength\abovedisplayskip{5pt}
		\setlength\belowdisplayskip{5pt}
		\small
		\label{MRC_LB_rate}
		R^U_k \ge {\underline R}_k^{{\rm MRC},U} = \frac{T_{U}}{T}\log_2(1+\text{SINR}^{\text{MRC},U}_k),
	\end{equation}
	where \(\text{SINR}^{\text{MRC},U}_k\) is given by
	\begin{equation}	\label{MRC_UP_rate}
		\text{SINR}^{\text{MRC},U}_k = \frac{M p^d_k  \times  \rho \tau p^p_k\beta^2_k |\Phi|^2 }{ (\sum\limits_{k'=1}^K p^d_{k'} \beta_{k'} + \frac{\sigma^2}{|\Phi|^2\rho})(\rho \tau p^p_k \beta_k |\Phi|^2 + \sigma^2) }.
	\end{equation}
	
	\emph{Proof}: Refer to Appendix \ref{Prooftheorem1}. $\hfill\blacksquare$
\end{theorem}

\begin{theorem}
	\label{ZF_SINR_T}
	The ergodic achievable rate for the $k$-th device using the ZF decoder can be lower bounded by
	\begin{equation}
		\setlength\abovedisplayskip{5pt}
		\setlength\belowdisplayskip{5pt}
		\small
		\label{ZF_LB_rate}
		R^{U}_k \ge {\underline R}_k^{{\rm ZF},U} = \frac{T_U}{T}\log_2(1+\text{SINR}^{\text{ZF},U}_k),
	\end{equation}
	where \(\text{SINR}^{\text{ZF},U}_k\) is given by
	\begin{equation}
		\label{ZF_up_SINR}
		\text{SINR}^{\text{ZF},U}_k = \frac{(M-K) p^d_k \times \rho \tau p^p_k\beta^2_k|\Phi|^2}{( \sum\limits_{k' = 1}^K p^d_{k'}e_{k'}  + \frac{\sigma^2}{|\Phi|^2 \rho})(\rho \tau p^p_k\beta_k|\Phi|^2 + \sigma^2)}.
	\end{equation}
	\emph{Proof}: Refer to Appendix \ref{Prooftheorem2}. $\hfill\blacksquare$
\end{theorem}

It is worth noting that the lower bounds for a conventional RF receiver can be obtained by substituting \(\Phi = 1\), \(\rho = \rho_{\text{RF}}\), and \(\sigma^2 = \sigma^2_{\text{RF}}\) into (\ref{MRC_UP_rate}) and (\ref{ZF_up_SINR}). Furthermore, under the harvested-energy constraint, the superiority of RAQRs over RF MIMO at limited transmit power becomes particularly evident.

\begin{corollary}\label{gain3}
	\emph{In the low-SINR regime (\(p^d_k\rightarrow0, p^p_k\rightarrow0\)), the rate difference between RAQRs and conventional MIMO is}
	\begin{equation}
		\begin{split}
			\Delta R^{\mathrm{MRC}/\mathrm{ZF},U}_k&\approx \frac{T_U}{T}\log_2\Big(1+\mathrm{SINR}_{k,\text{RF}}^{\mathrm{MRC}/\mathrm{ZF}}\frac{\rho^2|\Phi|^4}{\sigma^4}\Big),
		\end{split}
	\end{equation}
	where \(\mathrm{SINR}_{k,\text{RF}}^{\mathrm{MRC}/\mathrm{ZF}}\) is the SINR for MRC or ZF based on RF MIMO.
	
	\emph{Proof}: By using \(\log_2(1+x)\approx x/\ln(2)\) when \(x \rightarrow0\), we complete this proof. $\hfill\blacksquare$
\end{corollary}

\begin{remark}
	\emph{Compared to conventional RF MIMO, RAQRs exhibit a noticeable gain related to SINR in the low-power regime due to the ``squaring effect". }
\end{remark}

As stated in Corollary~\ref{gain3}, the performance gain originates from the normalized noise background \(\frac{\sigma^2}{\rho |\Phi|^2}\) inherent to Rydberg atoms. Consequently, part of the squaring gain is attributed to the reduced background noise, while the remaining part stems from improved channel estimation. Based on this observation, even though the harvest energy is constrained, the BS with RAQR is capable of capturing the weak RF signals from devices, thereby enabling battery-free communication.

%\begin{remark}\label{rem:RAQR-vs-trad}
%	\emph{Benefits of RAQR comes from two \textbf{power-free} settings ($\Phi,\rho$) and noise background. Furthermore, the gain of RAQR can be maximized by configuring \(\Phi\), confirming Remark 3 in \cite{gong2025rydberguplink}.}
	
%	As revealed in Corollary \ref{gain} and Corollary \ref{gain3}, the performance gain arises from the amplified gain and low background noise of Rydberg atoms. Compared with conventional receivers, RAQR introduces two controllable degrees of freedom, including
%	$\Phi$ and $\rho$.
%	In \eqref{eq:MRC-factor}–\eqref{eq:ZF-factor}, both MRC and ZF gain through \(\rho \Phi^2\), and thus
	%(i) \emph{without increasing pilot/transmission power}, RAQR can enhance the accuracy of channel estimation, thereby improving the system performance;
	%(ii) RAQR can also enhance the perfect-CSI baselines by decreasing the normalized noise background $\sigma^2/(\Phi^2\rho)$ in both MRC and ZF.
%\end{remark}

\subsection{Closed-Expression of SWIPT}
Assuming that the BS adopts the maximum rate transmission (MRT) and ZF proceding schemes, we have
\begin{equation}
	\mathbf{w}_k = \begin{cases}
	\frac{\mathbf{ \hat h}_k}{\mathbb{E}\Big\{\|\mathbf{ \hat h}_k\|^2\Big\}}, & \text{for } \text{MRT},\\
		\frac{\mathbf{\hat H}(\mathbf{\hat H}^H\mathbf{\hat H})^{-1}\boldsymbol{\epsilon}_k }{\mathbb{E}\Big\{\|\mathbf{\hat H}(\mathbf{\hat H}^H\mathbf{\hat H})^{-1}\boldsymbol{\epsilon}_k\|^2\Big\}}, & \text{for } \text{ZF},
	\end{cases}
\end{equation}
where \(\mathbf{\hat H} = [\mathbf{ \hat h}_1,\cdots,\mathbf{\hat h}_K]\) collects the estimated channels and \(\boldsymbol{\epsilon}_k\) denotes the \(k\)-th column of matrix \(\mathbf{I}_K\). Therefore, we have the following  theorems.
\begin{theorem}
	\label{MRCSWIPT}
	For the MRT precoding case, the harvested energy and downlink rate of the $k$-th device can be lower bounded by
	\begin{equation}
		\setlength\abovedisplayskip{5pt}
		\setlength\belowdisplayskip{5pt}
		\small
		\label{MRC_SWIPT}
		\begin{split}
			E_k \ge &{\underline E}^{\rm MRT}_k\\
			=& \frac{T_D}{T}\eta_{\text{EH}}(1-\alpha_k)\Big[\rho_{\text{RF}}p^s_kM(\beta_k-e_k) + \beta_k\sum_{k' = 1}^{K}\rho_{\text{RF}}p^s_{k'} \Big],
		\end{split}
	\end{equation}
	and 
	\begin{equation}
		\setlength\abovedisplayskip{5pt}
		\setlength\belowdisplayskip{5pt}
		\small
		\label{MRC_downlink_rate}
		\begin{split}
			&R^{D}_k \ge {\underline R}^{{\rm MRT},D}_k = \frac{T_{D}}{T}\times \\
			& \log_2\Big(1 + \frac{\rho_{\text{RF}}p^s_k\alpha_kM(\beta_k - e_k)}{\alpha_k\beta_{k}\sum_{k'=1}^{K}\rho_{\text{RF}}p^s_{k'} +\alpha_k\sigma^2_{\text{RF},k} + \sigma^2_{k,s} }\Big).
		\end{split}
	\end{equation}
	
	\emph{Proof}: The proof is omitted owing to similar process in Appendix \ref{Prooftheorem1}. $\hfill\blacksquare$
\end{theorem}

\begin{theorem}
	\label{ZFSWIPT}
	The harvested energy and downlink rate of the $k$-th device based on the ZF precoding scheme can be respectively lower bounded by
	\begin{equation}
		\setlength\abovedisplayskip{5pt}
		\setlength\belowdisplayskip{5pt}
		\small
		\label{ZF_SWIPT}
		\begin{split}
			E_k &\ge {\underline E}^{\rm ZF}_k 
		=  \frac{T_D}{T}\eta_{\text{EH}}(1-\alpha_k)\\
		&\times\Big[\rho_{\text{RF}}p^s_k(M-K)(\beta_k-e_k) + \sum_{k' = 1}^{K}\rho_{\text{RF}}p^s_{k'}e_k\Big].
		\end{split}
	\end{equation}
	and 
	\begin{equation}
		\setlength\abovedisplayskip{5pt}
		\setlength\belowdisplayskip{5pt}
		\small
		\label{ZF_downlink_rate}
		\begin{split}
			&R^{D}_k \ge {\underline R}^{{\rm ZF},D}_k = \frac{T_D}{T}\times \\
			 & \log_2\Big(1 + \frac{\rho_{\text{RF}}p^s_k\alpha_k(M-K)(\beta_k - e_k)}{\alpha_ke_k\sum_{k'=1}^{K}\rho_{\text{RF}}p^s_{k'} + \alpha_k\sigma^2_{\text{RF},k} + \sigma^2_{k,s} }\Big),
		\end{split}
	\end{equation}
	
	\emph{Proof}: The proof is omitted owing to similar process in Appendix \ref{Prooftheorem2}. $\hfill\blacksquare$
\end{theorem}

As seen from \eqref{MRC_downlink_rate} and \eqref{ZF_downlink_rate}, both the achievable rate and the harvested energy depend on the channel-estimation error \(e_k\), because the accuracy of the estimate directly determines the beam-alignment mismatch. Consequently, improved channel estimation for RAQR substantially enhances the efficiency of downlink SWIPT.

\section{Joint Design for Payload, Pilot and SWIPT for MRC/MRT}
In this section, we solve Problem (\ref{Optimization}) based on the derived lower bounds and then propose the joint design for uplink training, uplink transmission, and downlink SWIPT.

\subsection{Joint Power and Coefficient Optimization}
With the given transmission block \(T_D\) and \(T_U\), Problem (\ref{Optimization}) can be written as
\begin{subequations}
	\setlength\abovedisplayskip{5pt}
	\setlength\belowdisplayskip{5pt}
	\label{Optimization1}
	\begin{align}
		\small
		\mathop {\max }\limits_{\substack{ \left\{ {p_k^d} \right\},\left\{p^p_{k}\right\}\\
				\left\{\alpha_k\right\},\left\{ {p_k^s} \right\}}} \quad & \sum\limits_{k=1}^K (R^{{\rm MRT},D}_k + R^{{\rm MRC},U}_k)\notag\\
		{\rm{s}}{\rm{.t}}{\rm{.}}\;\;\;\; & {\underline R}^{{\rm MRC},U}_k \ge R_k^{{\rm{req}},U},\forall k, \label{Optimization1_A}\\
		& {\underline R}^{{\rm MRT},D}_k \ge R_k^{{\rm{req}},D},\forall k, \label{Optimization1_B}\\
		& \sum\limits_{k=1}^K p^s_k \le P^{s,\max},T \label{Optimization1_C} \\
	& K p^p_k + T_U p^d_k \le T{\underline E}^{\rm MRC}_k, \forall k, \label{Optimization1_D} \\
			&0 \le \alpha_k \le 1, \forall k. \label{Optimization1_E}
	\end{align}
\end{subequations}
However, it is challenging to solve this problem. To address this issue, Lemma 3 in \cite{peng2023resource} is adopted, which can be expressed as
\begin{equation}
	\log_2(1+x) \ge \zeta \log_2 x + \nu, (x \ge 0),
\end{equation}
where \(\zeta\) and \(\nu\) are given by
\begin{equation}\label{logappr}
	\zeta = \frac{\hat x}{1+\hat x}, \nu = \log_2(1+\hat x) - \zeta \log_2(\hat x).
\end{equation} 

By using this lemme, lower bound based on MRC can be approximated in an iterative manner. Then, we introduce the auxiliary variables \(\{\chi^D_k\}\) and \(\{\chi^U_k\}\) for downlink and uplink SINR, respectively. By ignoring the constant terms, Problem~\eqref{Optimization1} in the \(i\)-th iteration can be reformulated as
\begin{subequations}
	\setlength\abovedisplayskip{5pt}
	\setlength\belowdisplayskip{5pt}
	\label{Optimization1a}
	\begin{align}
		\small
		\mathop {\max }\limits_{\substack{ \left\{ {p_k^d} \right\},\left\{p^p_{k}\right\}, \left\{\alpha_k\right\}\\
				\left\{ {p_k^s} \right\},\left\{ {\chi^U_k} \right\} ,\left\{ {\chi^D_k} \right\}}}  \quad & \prod\limits_{k=1}^K (\chi^D_k)^{\zeta^{D,(i)}_k}(\chi^U_k)^{\zeta^{U,(i)}_k} \notag\\
		{\rm{s}}{\rm{.t}}{\rm{.}}\;\;\;\; & \chi^U_k \le \text{SINR}^{{\rm MRC},U}_k,\forall k, \label{Optimization1a_A}\\
		& \chi^D_k \le \text{SINR}^{{\rm MRC},D}_k,\forall k, \label{Optimization1a_B}\\
		& \chi^U_k \ge 2^\frac{TR_k^{{\rm{req}},U}}{T_U}-1, \forall k, \label{Optimization1a_C}\\
		& \chi^D_k \ge 2^\frac{TR_k^{{\rm{req}},D}}{T_D}-1, \forall k, \label{Optimization1a_D}\\
		& (\ref{Optimization1_C}),(\ref{Optimization1_D}),(\ref{Optimization1_E}), \label{Optimization1a_E}
	\end{align}
\end{subequations}
where \(\zeta^{U,(i)}_k\) and \(\zeta^{D,(i)}_k\) are obtained by~\eqref{logappr} in the \(i\)-th iteration. Note that Problem~\eqref{Optimization1a} cannot be readily solved using CVX, as constraint~\eqref{Optimization1_D} is not a monomial function. 

To address this issue, constraint~\eqref{Optimization1_D} can be rewritten as
\begin{equation}
	\begin{split}
			% &T_D\eta_{\text{EH}}\alpha_k\rho_{\text{RF}}\beta_k\Big(p^s_kM\frac{\rho\tau p^p_k\beta_{k}\Phi^2}{\rho\tau p^p_k\beta_k\Phi^2 + \sigma^2} + \sum_{k'=1}^{K}p^s_{k'}\Big)\\
			 % \ge& (K p^p_k + T^U p^d_k) \\
			  &M\rho \tau \beta_{k}|\Phi|^2 p^p_k p^s_k+\sum_{k'=1}^{K}p^s_{k'}(\rho\tau p^p_k\beta_k|\Phi|^2 + \sigma^2)\\
			  \ge&  \frac{(K p^p_k + T^U p^d_k) (\rho\tau p^p_k\beta_k|\Phi|^2 + \sigma^2)}{T_D\eta_{\text{EH}}\rho_{\text{RF}}\beta_k} +\alpha_kM\rho \tau \beta_{k}|\Phi|^2 p^p_k p^s_k\\
			  +&\alpha_k\sum_{k'=1}^{K}p^s_{k'}(\rho\tau p^p_k\beta_k|\Phi|^2 + \sigma^2).
	\end{split}
\end{equation}
Then, the following theorem is introduced.
\begin{theorem}\label{MRCapprox}
	For any given vector \(\mathbf{\hat p}^s = [{\hat p}^s_1,\cdots,{\hat p}^s_K]^T\) with \({\hat p}^s_{k'} \ge 0\), \(\forall k'\), \(A_{k'} > 0\), \(B_k > 0\), and \( {\hat p}^p_k \ge 0\),  \(\sum\limits_{k' = 1}^Kp^s_{k'} + p^p_k\sum\limits_{k' = 1}^KA_{k'}p^s_{k'}\) is lower bounded by
	\begin{equation}
		B_k\sum\limits_{k' = 1}^Kp^s_{k'} + p^p_k\sum\limits_{k' = 1}^KA_{k'}p^s_{k'} \ge \delta_k (p^p_k)^{\omega^p_{k}}\prod_{k' = 1}^{K} (p^s_{k'})^{\omega^s_{k,k'}},
	\end{equation}
	where \(\delta_k\), \({\omega^p_{k}}\) and \({\omega^s_{k,k'}}\), \(\forall k'\), are given by
		\begin{equation}\label{xi}
		\begin{split}
			\delta_k = \frac{B_k\sum\limits_{k' = 1}^K{\hat p}^s_{k'} + {\hat p}^p_k\sum\limits_{k' = 1}^KA_{k'}{\hat p}^s_{k'} }{({\hat p}^p_k)^{\omega^p_{k}}\prod_{k' = 1}^{K} ({\hat p}^s_{k'})^{\omega^s_{k,k'}}},
		\end{split}
	\end{equation} 
	\begin{equation}\label{omegap}
		\begin{split}
		 \omega^p_k = \frac{{\hat p}^p_k\sum\limits_{k' = 1}^KA_{k'}{\hat p}^s_{k'}}{	B_k\sum\limits_{k' = 1}^K{\hat p}^s_{k'} + {\hat p}^p_k\sum\limits_{k' = 1}^KA_{k'}{\hat p}^s_{k'}},
		\end{split}
	\end{equation} 
	and 
		\begin{equation}
		\begin{split}\label{omegas}
			\omega^s_{k,k'} = \frac{B_k{\hat p}^s_k + {\hat p}^p_k A_{k'}p^s_{k'}}{	B_k\sum\limits_{j = 1}^K{\hat p}^s_{j} + {\hat p}^p_k\sum\limits_{j = 1}^KA_{j}{\hat p}^s_{j}}, \forall k'.
		\end{split}
	\end{equation}
Note the the equality only holds at $(\hat{\mathbf p}^s,\hat p_k^p)$.

	\emph{Proof}: Please refer to Appendix~\ref{ProofMRCappro}. $\hfill\blacksquare$
\end{theorem}

By using Theorem~\eqref{MRCapprox}, constraint~\eqref{Optimization1_D} in the \(i\)-th iteration can be given by
\begin{equation}\label{ixi}
	\begin{split}
		&	\delta^{(i)}_k (p^p_k)^{\omega^{p,{(i)}}_{k}}\prod_{k' = 1}^{K} (p^s_{k'})^{\omega^{s,{(i)}}_{k,k'}}\ge \alpha_k\sum_{k'=1}^{K}p^s_{k'}(\rho\tau p^p_k\beta_k|\Phi|^2 + \sigma^2)\\
		+&   \frac{(K p^p_k + T^U p^d_k) (\rho\tau p^p_k\beta_k|\Phi|^2 + \sigma^2)}{T_D\eta_{\text{EH}}\rho_{\text{RF}}\beta_k} +\alpha_kM\rho \tau \beta_{k}|\Phi|^2 p^p_k p^s_k,
	\end{split}
\end{equation}
where 	\(\delta^{(i)}_k\), \({\omega^{p,{(i)}}_{k}}\) and \({\omega^{s,{(i)}}_{k'}}\) are obtained by using \eqref{xi}, \eqref{omegap}, and \eqref{omegas} in the \(i\)-th iteration, respectively. Finally, by substituting \eqref{ixi} into \eqref{Optimization1a}, Problem \eqref{Optimization1} can be reformulated as
\begin{subequations}
	\setlength\abovedisplayskip{5pt}
	\setlength\belowdisplayskip{5pt}
	\label{Optimization1b}
	\begin{align}
		\small
		\mathop {\max }\limits_{\substack{ \left\{ {p_k^d} \right\},\left\{p^p_{k}\right\}, \left\{\alpha_k\right\}\\
				\left\{ {p_k^s} \right\},\left\{ {\chi^U_k} \right\} ,\left\{ {\chi^D_k} \right\}}} \quad & \prod\limits_{k=1}^K (\chi^D_k)^{\zeta^{D,(i)}_k}(\chi^U_k)^{\zeta^{U,(i)}_k} \notag\\
		{\rm{s}}{\rm{.t}}{\rm{.}}\;\;\;\; & \eqref{Optimization1a_A},\eqref{Optimization1a_B},\eqref{Optimization1a_C},\eqref{Optimization1a_D},\eqref{Optimization1a_E},\eqref{ixi},
	\end{align}
\end{subequations}
It can be readily solved by using CVX, which is detailed in Algorithm \ref{alg:MRC}.

\begin{algorithm}[t]
	\caption{Iterative Algorithm for Problem \eqref{Optimization1} under MRC}
	\label{alg:MRC}
	\begin{algorithmic}[1]
	\STATE Initialize iteration number $i = 1$, and error tolerance $\kappa = 0.01$; 
	\STATE Initialize the power $ \left\{p^{p,\left( 1\right)}_k,p^{d,\left( 1\right)}_k,p^{s,\left( 1\right)}_k,\forall k \right\}$ and power-splitting coefficient \(\{\alpha_k, \forall k\}\), calculate SINR $\left \{\chi _k^{U,\left( 1 \right)},\chi _k^{D,\left( 1 \right)},\forall k\right\}$, obtain the sum rate and denoted as ${\rm{Obj}}^{\left(1\right)}$. Set ${\rm{Obj}}^{\left(0\right)} = 0$; 
	\WHILE {$ {\rm{Obj}}^{\left(i\right)}-{\rm{Obj}}^{\left(i-1\right)} \ge \kappa{{\rm{Ob}}{{\rm{j}}^{\left( {i - 1} \right)}}}$} 
	\STATE Update $\left \{ \zeta^{U,(i)}_k,\zeta^{U,(i)}_k, \delta^{(i)}_k, \omega^{p,(i)}_k,\omega^{s,(i)}_{k,k},\forall k \right\}$; 
	\STATE Update $i = i+1$, solve Problem (\ref{Optimization1b}) by using the CVX package to obtain $\left \{ p_k^{p,\left( i \right)},p_k^{d,\left( i \right)},p_k^{s,\left( i \right)},\alpha^{(i)}_k,\forall k\right\}$, calculate SINR $\left \{\chi _k^{U,\left( i \right)},\chi _k^{D,\left( i \right)},\forall k\right\}$ and then obtain the weighted sum rate, denoted as ${\rm{Obj}}^{\left(i\right)}$; 
	\ENDWHILE
	\end{algorithmic}
\end{algorithm}

% the best local monomial approximation

\subsection{Block Optimization}
With the fixed power \(\{p^p_k,p^d_k,p^s_k,\forall k\}\) and coefficient \(\alpha_k, \forall k\), Problem \eqref{Optimization} can be written as
\begin{subequations}
	\setlength\abovedisplayskip{5pt}
	\setlength\belowdisplayskip{5pt}
	\label{Optimization3}
	\begin{align}
		\small
		\mathop {\max }\limits_{\left\{ T_U\right\},\left\{ T_D\right\}} \quad & \sum\limits_{k=1}^K (R^{{\rm MRT},D}_k + R^{{\rm MRC},U}_k)\notag\\
		{\rm{s}}{\rm{.t}}{\rm{.}}\;\;\;\; & \eqref{Optimization_A},\eqref{Optimization_B}, \eqref{Optimization_D},\eqref{Optimization_F}, \eqref{Optimization_G}.
	\end{align}
\end{subequations}
The problem is a mixed-integer program. To tackle it, we relax the integer constraint \eqref{Optimization_G} and obtain a linear program, which can be efficiently solved by CVX. Finally, the transmission-block lengths are recovered by rounding the relaxed solution to the nearest integers.

\subsection{Algorithm Analysis}
\subsubsection{Convergence}
As can be seen from Algorithm \ref{alg:MRC}, we first check whether the solutions in the \(i\)-th iteration are feasible or not in the \((i+1)\)-th iteration. In the \(i\)-th iteration, constraint \eqref{Optimization1_D} can be expressed as
\begin{equation}
		\begin{split}
		&	\delta^{(i-1)}_k (p^{p,(i)}_k)^{\omega^{p,{(i-1)}}_{k}}\prod_{k' = 1}^{K} (p^{s,(i)}_{k'})^{\omega^{s,{(i-1)}}_{k,k'}}\\
		\ge& \alpha^{(i)}_k\sum_{k'=1}^{K}p^{s,(i)}_{k'}(\rho\tau p^{p,(i)}_k\beta_k|\Phi|^2 + \sigma^2)\\
		+&   \frac{(K p^{p,(i)}_k + T^U p^{d,(i)}_k) (\rho\tau p^{p,(i)}_k\beta_k|\Phi|^2 + \sigma^2)}{T_D\eta_{\text{EH}}\rho_{\text{RF}}\beta_k} \\
		+&\alpha^{(i)}_kM\rho \tau \beta_{k}|\Phi|^2 p^{p,(i)}_k p^{s,(i)}_k,
	\end{split}
\end{equation}
where \(\left\{p^{p,(i)}_k,p^{s,(i)}_k,p^{d,(i)}_k,\alpha^{(i)}_k, \forall k\right\}\) are the optimal solution in the \(i\)-th iteration. By using Theorem \ref{MRCapprox}, we have
\begin{equation}
	\begin{split}
	&	M\rho \tau \beta_{k}|\Phi|^2 p^{p,(i)}_k p^{s,(i)}_k+\sum_{k'=1}^{K}p^{s,(i)}_{k'}(\rho\tau p^{p,(i)}_k\beta_k|\Phi|^2 + \sigma^2) \\
		\ge & 	\delta^{(i)}_k (p^{p,(i)}_k)^{\omega^{p,{(i)}}_{k}}\prod_{k' = 1}^{K} (p^{s,(i)}_{k'})^{\omega^{s,{(i)}}_{k,k'}}\\
		\ge & 	\delta^{(i-1)}_k (p^{p,(i)}_k)^{\omega^{p,{(i-1)}}_{k}}\prod_{k' = 1}^{K} (p^{s,(i)}_{k'})^{\omega^{s,{(i-1)}}_{k,k'}}.
	\end{split}
\end{equation}
As a result, we prove that the solutions are still feasible.

Similarly, we prove that the rate in the \(i\)-th iteration is larger than that of the \((i-1)\)-th iteration, which can be expressed as
\begin{equation}
	\begin{split}
		&\log_2(1+\text{SINR}^{{\rm MRC},U,(i)}_k) \\
		= &\zeta^{U/D,(i)}_k \log_2 (\text{SINR}^{{\rm MRC},U,(i)}_k)+\nu^{U,(i)}_k \\
		\ge& \zeta^{U/D,(i-1)}_k \log_2 (\text{SINR}^{{\rm MRC},U,(i)}_k)+\nu^{U,(i-1)}_k\\
		\ge &\zeta^{U/D,(i-1)}_k \log_2 (\text{SINR}^{{\rm MRC},U,(i-1)}_k)+\nu^{U,(i-1)}_k \\
		=& \log_2(1+\text{SINR}^{{\rm MRC},U,(i-1)}_k).
	\end{split}
\end{equation}
Similarly, the convergence of downlink rate can be proved.
Based on the above inequality, we complete the convergence of Algorithm \ref{alg:MRC}.

Finally, for given transmit powers and the power-splitting coefficient, the optimal time allocation is obtained via a linear program. By denoting the optimal solution as \((T_U^{\star}, T_D^{\star})\), the sum rate is always no larger than
\begin{equation}
	\begin{split}
		\sum_{k=1}^{K} \big( R^{\mathrm{MRT},D}_k + R^{\mathrm{MRC},U}_k \big)\big|_{\,T_U^{\star},\,T_D^{\star}}.
	\end{split}
\end{equation}

As a result, the objective is non-decreasing across iterations and is upper bounded by the feasible rate region, which ensures convergence of the proposed algorithm.

\subsubsection{Complexity}
GP can be efficiently solved via standard interior-point methods with the worst-case polynomial-time complexity. For Algorithm~\ref{alg:MRC}, the computational complexity is on the order of ${\mathcal{O}}(N \times \max\{(6K)^{3}, N_{cost}\})$, where \(N\) is the number of iterations and $N_{cost}$ is the computational complexity of calculating the first-order and second-order derivatives of the objective function and constraint functions of Problem \eqref{Optimization1b} \cite{van2018joint}. In comparison, the complexity of the associated linear program is negligible relative to that of the GP.

\section{Joint Design for Payload, Pilot and SWIPT for ZF}
In this section, the power, power splitting coefficient, and block are devised for ZF detection/precoding.
\subsection{Joint Power and Coefficient Optimization}
Similarly, we introduce the auxiliary variables \(\{\chi^D_k,\chi^U_k,\forall k\}\) and solve it iteratively. Problem (\ref{Optimization}) based on ZF detection can be written as
\begin{subequations}
	\setlength\abovedisplayskip{5pt}
	\setlength\belowdisplayskip{5pt}
	\label{OptimizationZF}
	\begin{align}
		\small
		\mathop {\max }\limits_{\substack{ \left\{ {p_k^d} \right\},\left\{p^p_{k}\right\}, \left\{\alpha_k\right\}\\
				\left\{ {p_k^s} \right\},\left\{ {\chi^U_k} \right\} ,\left\{ {\chi^D_k} \right\}}}  \quad & \prod\limits_{k=1}^K (\chi^D_k)^{\zeta^{D,(i)}_k}(\chi^U_k)^{\zeta^{U,(i)}_k} \notag\\
		{\rm{s}}{\rm{.t}}{\rm{.}}\;\;\;\; & \chi^U_k \le \text{SINR}^{{\rm ZF},U}_k,\forall k, \label{OptimizationZF_A}\\
		& \chi^D_k \le \text{SINR}^{{\rm ZF},D}_k,\forall k, \label{OptimizationZF_B}\\
		& \chi^U_k \ge 2^\frac{TR_k^{{\rm{req}},U}}{T_U}-1, \forall k, \label{OptimizationZF_C}\\
		& \chi^D_k \ge 2^\frac{TR_k^{{\rm{req}},D}}{T_D}-1, \forall k, \label{OptimizationZF_D}\\
			& K p^p_k + T_U p^d_k \le T{\underline E}^{\rm ZF}_k , \forall k, \label{OptimizationZF_E}\\
		& (\ref{Optimization_C}),(\ref{Optimization_F}),(\ref{Optimization_E}). \label{OptimizationZF_F}
	\end{align}
\end{subequations}
However, Problem~\eqref{OptimizationZF} cannot be directly solved owing to constraints (\ref{OptimizationZF_A}) and \eqref{OptimizationZF_E}. To make it more clear, we rewrite~\eqref{ZF_up_SINR} as \eqref{reZF}, which is given at the bottom of this page.
\begin{figure*}[b]
	\hrule
	\begin{equation}\label{reZF}
		\begin{split}
			\text{SINR}^{{\rm ZF},U}_k 
			& =  \frac{(M-K) p^d_k \times \rho \tau p^p_k\beta^2_k|\Phi|^2 \times \prod\limits_{j=1}^{K}(1+\frac{\rho\tau p^p_j\beta_j|\Phi|^2}{\sigma^2})}{ \Big[ \sum\limits_{k' = 1}^Kp^d_{k'}\beta_{k'}\prod\limits_{j\neq k'}^{K}(1+\frac{\rho\tau p^p_j\beta_j|\Phi|^2}{\sigma^2})  + \frac{\sigma^2}{\rho|\Phi|^2}\prod\limits_{j=1}^{K}(1+\frac{\rho\tau p^p_j\beta_j|\Phi|^2}{\sigma^2})\Big]( \rho\tau p^p_k\beta_k|\Phi|^2 + \sigma^2)}
		\end{split}
	\end{equation}
\end{figure*}
To tackle this issue, we adopt the following Theorem.
\begin{theorem}\label{ZFapprox}
	For any given vector \(\mathbf{\hat x} = [{\hat x}_1,\cdots,{\hat x}_K]^T\), with \({\hat x}_k \ge 0\), \(\forall k\), \(\prod_{k=1}^{K}(1 + x_k)\) is lower bounded by
	\begin{equation}
		\prod_{k=1}^{K}(1 + x_k) \ge \psi \prod_{k = 1}^{K}x^{\varphi_k}_k,
	\end{equation}
where	\(\psi\) and \(\varphi_k\), \(\forall k\) are given by
	\begin{equation}\label{varphi}
		\begin{split}
			\psi = \frac{\prod_{k=1}^{K}(1+\hat{x}_k)}{\prod_{k=1}^{K}\hat{x}^{\varphi_k}_k}, \varphi_k = \frac{\hat{x}_k}{1+\hat{x}_k}.
		\end{split}
	\end{equation} 
	Note that the equality only holds  when \(x_k = \hat{x}_k\), \(\forall k\).
	
	\emph{Proof}: Refer to Appendix E in \cite{ren2020joint}. $\hfill\blacksquare$
\end{theorem}

By using Theorem \ref{ZFapprox}, \(\prod\limits_{j=1}^{K}(1+\frac{\rho\tau p^p_j\beta_j|\Phi|^2}{\sigma^2})\) can be approximated by the best local monomial approximations in the \(i\)-th iteration, which can be expressed as
\begin{equation}\label{zfith}
	\begin{split}
		\prod\limits_{j=1}^{K}(1+\frac{\rho\tau p^p_j\beta_j|\Phi|^2}{\sigma^2}) \ge  \psi^{(i)} \prod_{k = 1}^{K}(\frac{\rho\tau p^p_j\beta_j|\Phi|^2}{\sigma^2})^{\varphi^{(i)} _j},
	\end{split}
\end{equation}
where \(\psi^{(i)}\) and \({\varphi^{(i)} _j}\) are obtained by \eqref{varphi} in the \(i\)-th iteration. Substituting \eqref{zfith} into constraints (\ref{OptimizationZF_A}), we have
\begin{equation}\label{rateappro}
	\begin{split}
	 &\psi^{(i)} \prod_{k = 1}^{K}(\frac{\rho\tau p^p_j\beta_j|\Phi|^2}{\sigma^2})^{\varphi^{(i)} _j} (M-K) p^d_k \times \rho \tau p^p_k\beta^2_k|\Phi|^2 \\
	 \ge & \Big\{ \sum\limits_{k' = 1}^Kp^d_{k'}\beta_{k'}\prod\limits_{j\neq k'}^{K}(1+\frac{\rho\tau p^p_j\beta_j|\Phi|^2}{\sigma^2})  
	 + \frac{\sigma^2}{\rho|\Phi|^2}
	 \\
	 \times &\prod\limits_{j=1}^{K}(1+\frac{\rho\tau p^p_j\beta_j|\Phi|^2}{\sigma^2})\Big\} ( \rho\tau p^p_k\beta_k|\Phi|^2 + \sigma^2).
	\end{split}
\end{equation}

Based on Theorem \ref{MRCapprox}, constraints \eqref{OptimizationZF_E} can be approximated in an iterative manner, which can be written as
\begin{equation}\label{Ekappro}
	\begin{split}
		  &(M - K)\rho \tau \beta_{k}|\Phi|^2 p^p_k p^s_k+\sigma^2\sum_{k'=1}^{K}p^s_{k'}\\
		  \ge & \delta^{{\rm ZF},(i)}_k (p^p_k)^{\omega^{{\rm ZF},p,{(i)}}_{k}}\prod_{k' = 1}^{K} (p^s_{k'})^{\omega^{{\rm ZF},s,{(i)}}_{k,k'}} \\
		\ge&  \frac{(K p^p_k + T^U p^d_k) (\rho\tau p^p_k\beta_k|\Phi|^2 + \sigma^2)}{T_D\eta_{\text{EH}}\rho_{\text{RF}}\beta_k}\\ +&\alpha_k(M - K)\rho \tau \beta_{k}|\Phi|^2 p^p_k p^s_k
		+\alpha_k\sigma^2\sum_{k'=1}^{K}p^s_{k'},
	\end{split}
\end{equation}
where \(\delta^{{\rm ZF},(i)}_k\), \({\omega^{{\rm ZF},p,{(i)}}_{k}}\), and \({\omega^{{\rm ZF},s,{(i)}}_{k,k'}} \) are obtained by using \eqref{xi}, \eqref{omegap}, and \eqref{omegas}, respectively.

Based on the above discussions, Problem \eqref{OptimizationZF} can be formulated as
\begin{subequations}
	\setlength\abovedisplayskip{5pt}
	\setlength\belowdisplayskip{5pt}
	\label{OptimizationZF1}
	\begin{align}
		\small
		\mathop {\max }\limits_{\substack{ \left\{ {p_k^d} \right\},\left\{p^p_{k}\right\}, \left\{\alpha_k\right\}\\
				\left\{ {p_k^s} \right\},\left\{ {\chi^U_k} \right\} ,\left\{ {\chi^D_k} \right\}}}  \quad & \prod\limits_{k=1}^K (\chi^D_k)^{\zeta^{D,(i)}_k}(\chi^U_k)^{\zeta^{U,(i)}_k} \notag\\
		{\rm{s}}{\rm{.t}}{\rm{.}}\;\;\;\; & \eqref{rateappro}, \eqref{OptimizationZF_B}, \eqref{OptimizationZF_C},\eqref{OptimizationZF_D},\eqref{Ekappro},\eqref{OptimizationZF_F}.
	\end{align}
\end{subequations}
As can be seen, it can be solved by using CVX, which is detailed in Algorithm \ref{alg:ZF}.

\begin{algorithm}[t]
	\caption{Iterative Algorithm for Problem \eqref{OptimizationZF} under ZF}
	\label{alg:ZF}
	\begin{algorithmic}[1]
		\STATE Initialize iteration number $i = 1$, and error tolerance $\kappa = 0.01$; 
		\STATE Initialize the power $ \left\{p^{p,\left( 1\right)}_k,p^{d,\left( 1\right)}_k,p^{s,\left( 1\right)}_k,\forall k \right\}$ and coefficient \(\{\alpha_k, \forall k\}\), calculate SINR $\left \{\chi _k^{U,\left( 1 \right)},\chi _k^{D,\left( 1 \right)},\forall k\right\}$, obtain the sum rate and denoted as ${\rm{Obj}}^{\left(1\right)}$. Set ${\rm{Obj}}^{\left(0\right)} = 0$; 
		\WHILE {$ {\rm{Obj}}^{\left(i\right)}-{\rm{Obj}}^{\left(i-1\right)} \ge \kappa{{\rm{Ob}}{{\rm{j}}^{\left( {i - 1} \right)}}}$} 
		\STATE Update $\left \{ \zeta^{U,(i)}_k,\zeta^{U,(i)}_k, \delta^{{\rm ZF},(i)}_k, \omega^{{\rm ZF},p,(i)}_k,\omega^{{\rm ZF},s,(i)}_{k,k},\varphi^{(i)}_k \right\}$ and \(\psi^{(i)}\); 
		\STATE Update $i = i+1$, solve Problem (\ref{Optimization1b}) by using the CVX package to obtain $\left \{ p_k^{p,\left( i \right)},p_k^{d,\left( i \right)},p_k^{s,\left( i \right)},\alpha^{(i)}_k,\forall k\right\}$, calculate SINR $\left \{\chi _k^{U,\left( i \right)},\chi _k^{D,\left( i \right)},\forall k\right\}$ and then obtain the weighted sum rate, denoted as ${\rm{Obj}}^{\left(i\right)}$; 
		\ENDWHILE
	\end{algorithmic}
\end{algorithm}

\subsection{Transmission Block Optimization}
Similar to the MRC case, when the transmit powers and the power-splitting coefficient are fixed, the transmission-block allocation can be solved by using a relaxation and a linear-program formulation, the derivation of which is omitted for brevity. The convergence guarantee and computational complexity follow the same procedure as for MRC and are not repeated here.

\section{Simulation Results}
In this section, we employ Monte Carlo simulations to validate the derived closed-form expressions and to investigate the impacts of channel estimation accuracy and the Rician factor on system performance. We then highlight the advantages of RAQ-MIMO over conventional RF MIMO and provide a comprehensive performance analysis.

\subsection{Parameter Settings}

\begin{table}[t]
	\renewcommand{\arraystretch}{1} % 行距微调，安全无依赖
	\caption{System Parameters}
	\label{tab:params}
	\centering
	\begin{tabular}{c l c }
		\hline
		\textbf{Symbol} & \textbf{Description} & \textbf{Value}  \\
		\hline
		$M$                 & Number of antennas (vapor cells)              & 100   \\
		$K$                 & Number of IoT devices                       & 10    \\
		$P^{s,\max}$               & Downlink (SWIPT) power                & 50 (W)   \\
		$\tau$            & Pilot length                          & 10     \\
		$T$                 & Coherence block length                & 400    \\
			$\eta_{\rm EH}$                 & Harvested energy efficiency               & 0.2    \\
			$R^{{\rm req},U}_k$                 & Uplink rate              & 0.2 bit/s/Hz  \\
				$R^{{\rm req},D}_k$                 & Downlink rate              & 1  bit/s/Hz  \\
		\hline
	\end{tabular}
\end{table}

We consider the four-level electron transition scheme of \(6{\rm S}_{1/2}\rightarrow6{\rm S}_{3/2}\rightarrow47{\rm D}_{5/2}\rightarrow48{\rm P}_{3/2}\). Unless otherwise stated, the quantum parameters are given in Table \(\textcolor{red}{\rm I}\) in \cite{gong2024rydberg}, and the remaining parameters are given in Table I. We assume that \(K=10\) devices are randomly distributed within a circular region of radius 50 meters, while their signals are collected by a BS with a RF transmitter and RAQRs located 150 meters from the center of the device region. The large-scale fading factors (dB) can be obtained by using \(-32.4-20\lg(d_k)-20\lg(f_c)\), where \(d_k\) (in meter) is the distance between the device \(k\) and RAQR, and \(f_c\) (in GHz) is the carrier frequency. The small-scale fading factors follow the Gaussian distribution of zero mean and unit variance. The noise background related to RF signal is assumed to be the same, i.e., \(\sigma^2_{\text{RF}} = \sigma^2_{\text{RF},k} = \sigma^2_{k,s}, \forall k\).  The power splitting coefficients \(\{\alpha_k, \forall \}\) are randomly chosen from (0,1).
Furthermore, the simulation results are averaged over \(10^{4}\) realizations.

\subsection{Channel Estimation}
\begin{figure}
	\centering
	\includegraphics[width=3in]{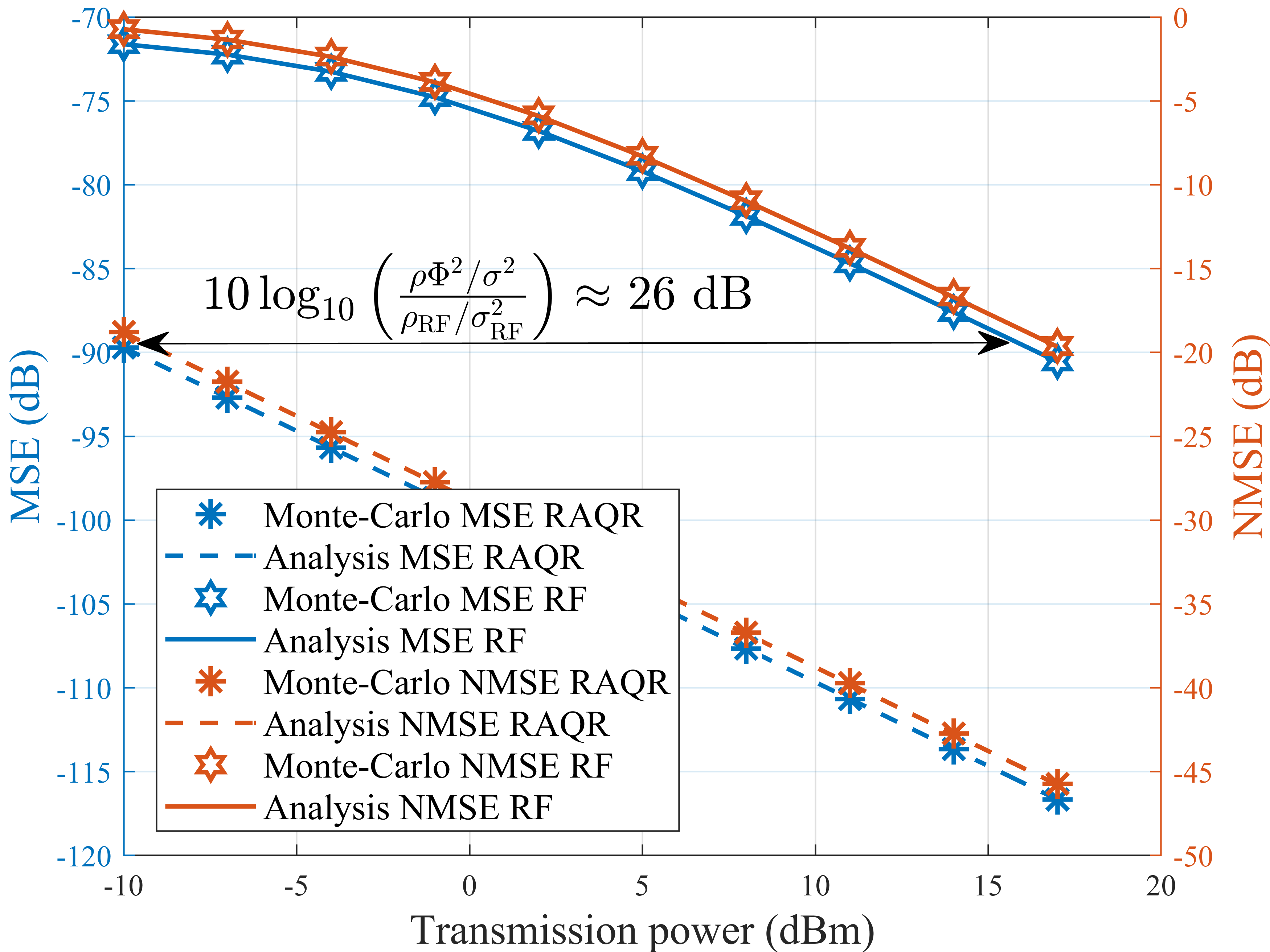}
	\caption{MSE under various pilot powers with \(M = 100\).}
	\label{MSE}
\end{figure}
We first evaluate the MSE and NMSE of RAQR with a fixed number of antennas (atomics) \(M = 100\) and compare the results with those of conventional MIMO, as illustrated in Fig. \ref{MSE}. As expected, both MSE and NMSE decrease as the pilot power increases, and the Monte Carlo simulation also validates the correctness of our derivation. Compared to conventional RF MIMO, it is noteworthy that the RAQR achieves superior performance owing to its enhanced normalized background noise \(\sigma^2/(\rho|\Phi|^2)\). Moreover, we observe that the RAQR achieves a power gain of approximately 26 dB over conventional MIMO, which validates the accuracy of Corollary 1 in Section II. This unique advantage enables RAQRs to obtain more accurate channel estimates even at low terminal transmit powers, thereby reducing uplink detection and downlink precoding errors.

\subsection{Lower Bounds}
\begin{figure*}[ht]
	\centering
	\subfigure[RAQR Uplink Rate with \(p^p_k = p^d_k,\forall k\).]{
		\begin{minipage}[t]{0.3\linewidth}
			\centering
			\includegraphics[width=2.3in]{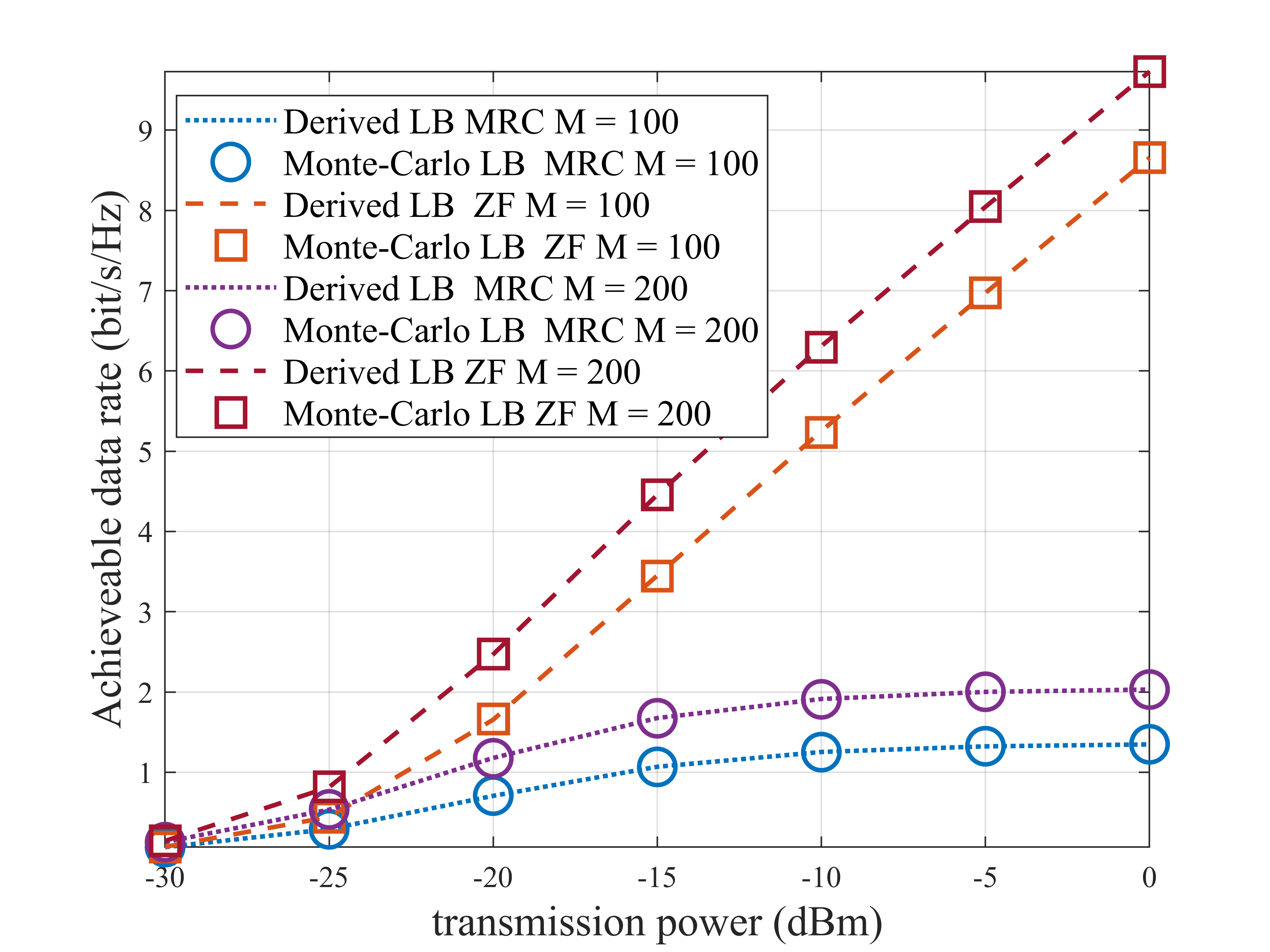}\hspace{10mm}
			%\caption{fig1}
	\end{minipage}}
	\quad%\includegraphics[width=2.75in]{cluster.png}}\hspace{10mm}
	\subfigure[RF Downlink Rate with  \(p^p_k = p^s_k,\forall k\).]{
	\begin{minipage}[t]{0.3\linewidth}
		\centering
		\includegraphics[width=2.3in]{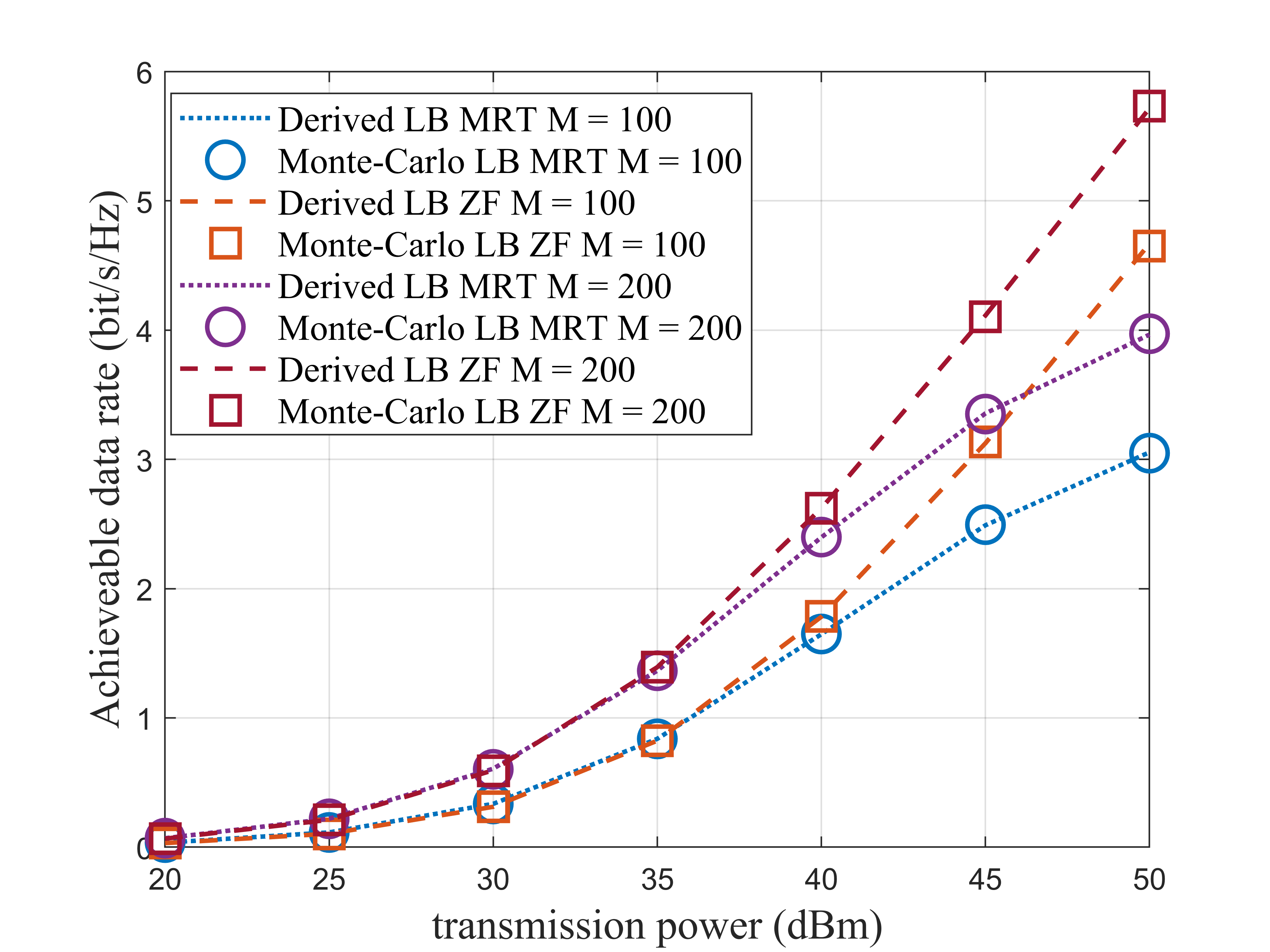}\hspace{10mm}
		%\caption{fig1}
\end{minipage}}
	\subfigure[Downlink Harvested Energy \(p^p_k = p^s_k,\forall k\).]{
	\begin{minipage}[t]{0.3\linewidth}
		\centering
		\includegraphics[width=2.3in]{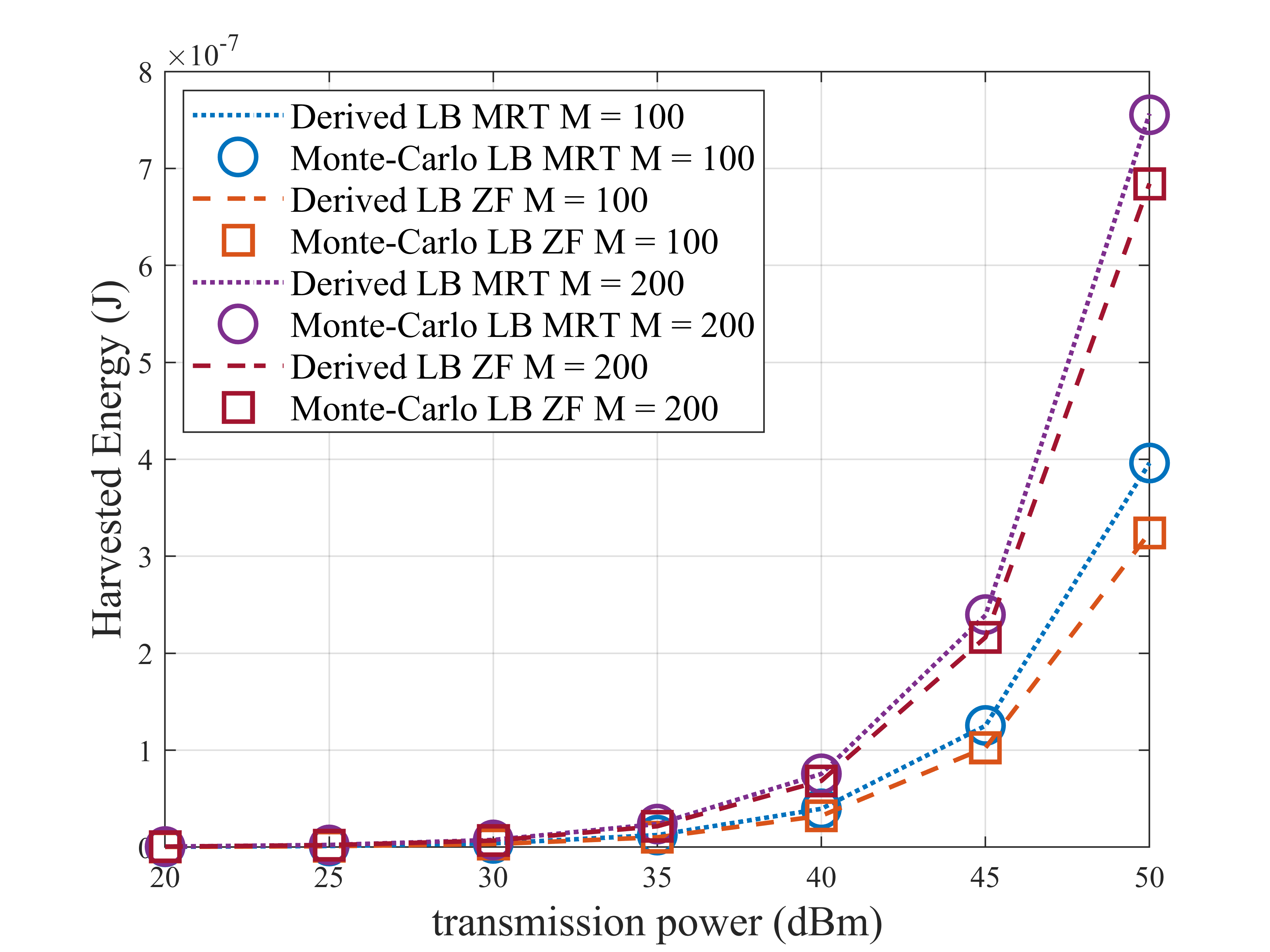}\hspace{10mm}
		%\caption{fig1}
\end{minipage}}
\caption{Derived lower bounds under various conditions.}
\label{lowerbounds}
\end{figure*}

In this subsection, we validate the tightness of the derived lower bounds by comparing them with Monte Carlo simulations in Fig.~\ref{lowerbounds}. As expected, the analytical expressions closely match the simulation results averaged over \(10^4\) trails. The achievable rate and the harvested energy both increase with the transmit power and the number of antennas. Furthermore, due to long-distance propagation, the harvested energy is relatively small, which leads to extremely low transmit power at IoT devices. Nevertheless, we observe that the uplink rate can approach the downlink rate even with minimal uplink transmit power, highlighting the unique advantages of RAQRs. This indicates that RAQR-enabled BSs can reliably receive weak RF signals from IoT devices, thereby enabling battery-free communication.

\subsection{Convergence}

\begin{figure}[h]
	\centering
	\includegraphics[width=3in]{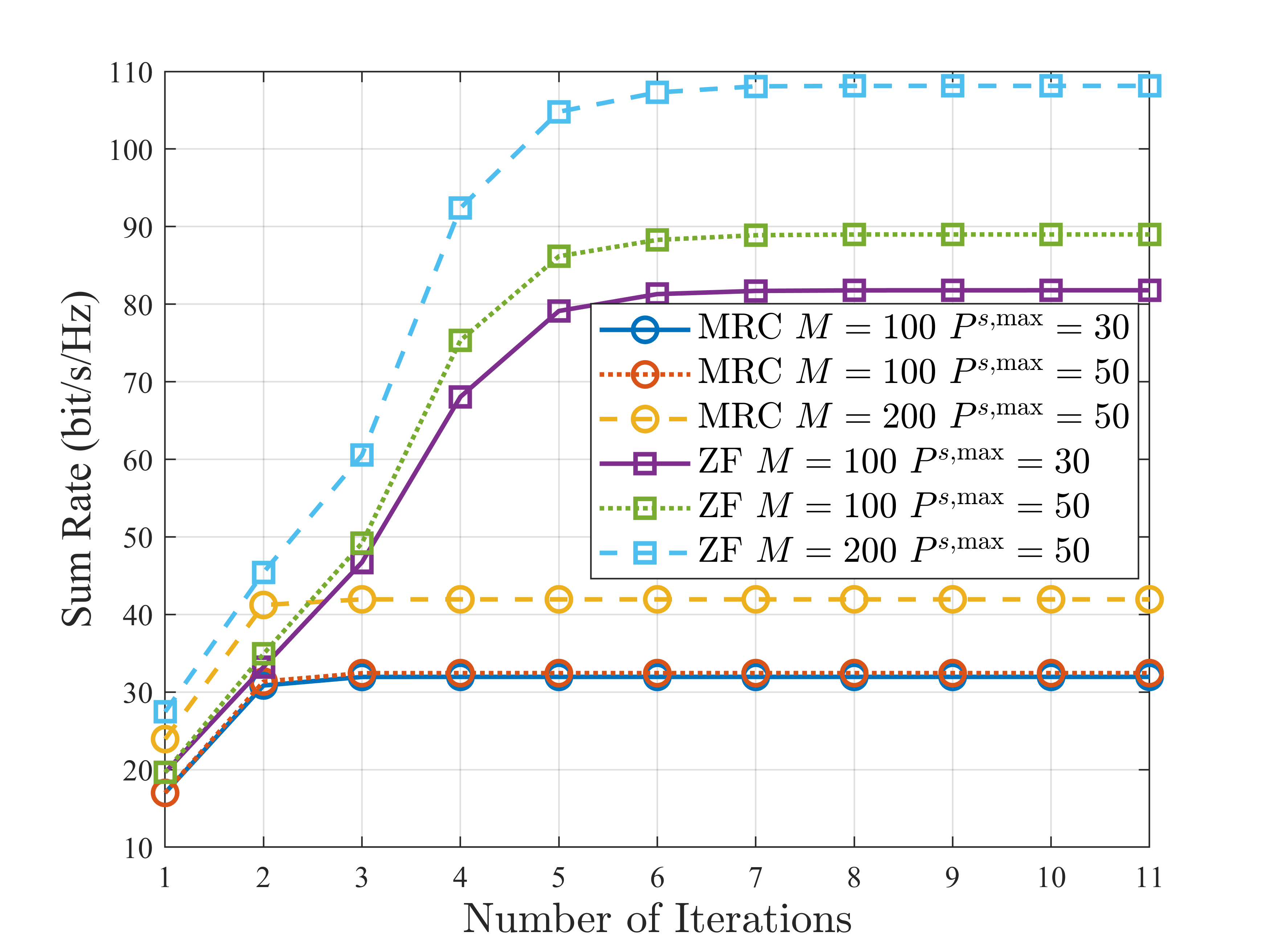}
	\caption{Convergence of proposed algorithm.}
	\label{convergence}
\end{figure}
In Fig.~\ref{convergence}, the convergence of our proposed algorithm is illustrated. We find that our proposed method converge to the locally optimal solution within 10 iterations, which verifies the effectiveness of our proposed algorithm. Furthermore, the sum rate increases with the increasing downlink power and number of antennas (vapor cells).

\subsection{Performance Evaluation}
When the BS employs a conventional RF receiver instead of an RAQR, reliably capturing weak RF signals becomes challenging. To sustain uplink transmission, IoT devices therefore require auxiliary batteries rather than relying solely on harvested energy. We refer to this baseline as \textbf{RF Receiver \& Battery-Powered Devices}. Furthermore, to evaluate the proposed methods, we consider the following benchmarks:
\begin{enumerate}
	\item \textbf{Benchmark~1 (RAQR, equal UL powers):}
	Under the RAQR-enabled setting, we enforce equal uplink pilot and payload powers, i.e., \(p_k^{p}=p_k^{d}\) for all \(k\). Given this constraint, the transmit power \(\{p_k^{d}, p^s_k\}\) and the power-splitting coefficients \(\{\alpha_k\}\) are optimized via GP.
	
	\item \textbf{Benchmark~2 (RF receiver \& battery-powered, full optimization):}
	We replace the RAQR with a conventional RF receiver and assume IoT devices are battery powered (\textbf{RF Receiver \& Battery-Powered Devices}). The uplink and downlink transmit powers together with power-splitting coefficients \(\{p_k^{p},p_k^{d}, p^s_k,\alpha_k\}\) are jointly optimized using the proposed algorithm.
	
	\item \textbf{Benchmark~3 (RF receiver \& battery-powered, equal UL powers):}
	Building on the RF baseline above, we additionally impose \(p_k^{p}=p_k^{d}\) for all \(k\). Then, the variables (e.g., transmission power \(\{p_k^{d}, p^s_k\}\) and \(\{\alpha_k\}\)) are optimized accordingly.
\end{enumerate}

\subsubsection{Effects of Power}
\begin{figure*}[ht]
	\centering
	\subfigure[MRC (MRT) with \(R^{{\rm req},U}_k = 1 \), \(R^{{\rm req},U}_k = 1\), \(\forall k\).]{
		\begin{minipage}[t]{0.45\linewidth}
			\centering
			\includegraphics[width=2.75in]{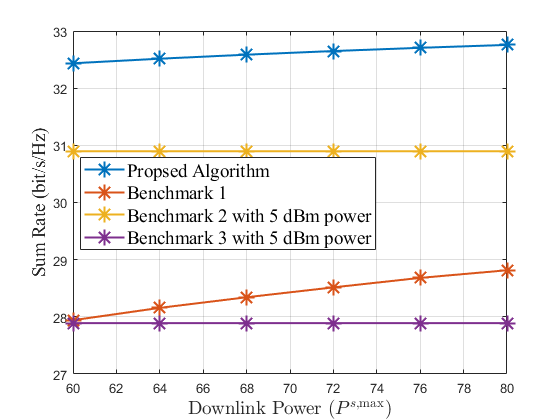}\hspace{10mm}
			%\caption{fig1}
	\end{minipage}}
	\quad%\includegraphics[width=2.75in]{cluster.png}}\hspace{10mm}
\subfigure[ZF with \(R^{{\rm req},U}_k = 1 \), \(R^{{\rm req},U}_k = 1\), \(\forall k\).]{
	\begin{minipage}[t]{0.45\linewidth}
		\centering
		\includegraphics[width=2.75in]{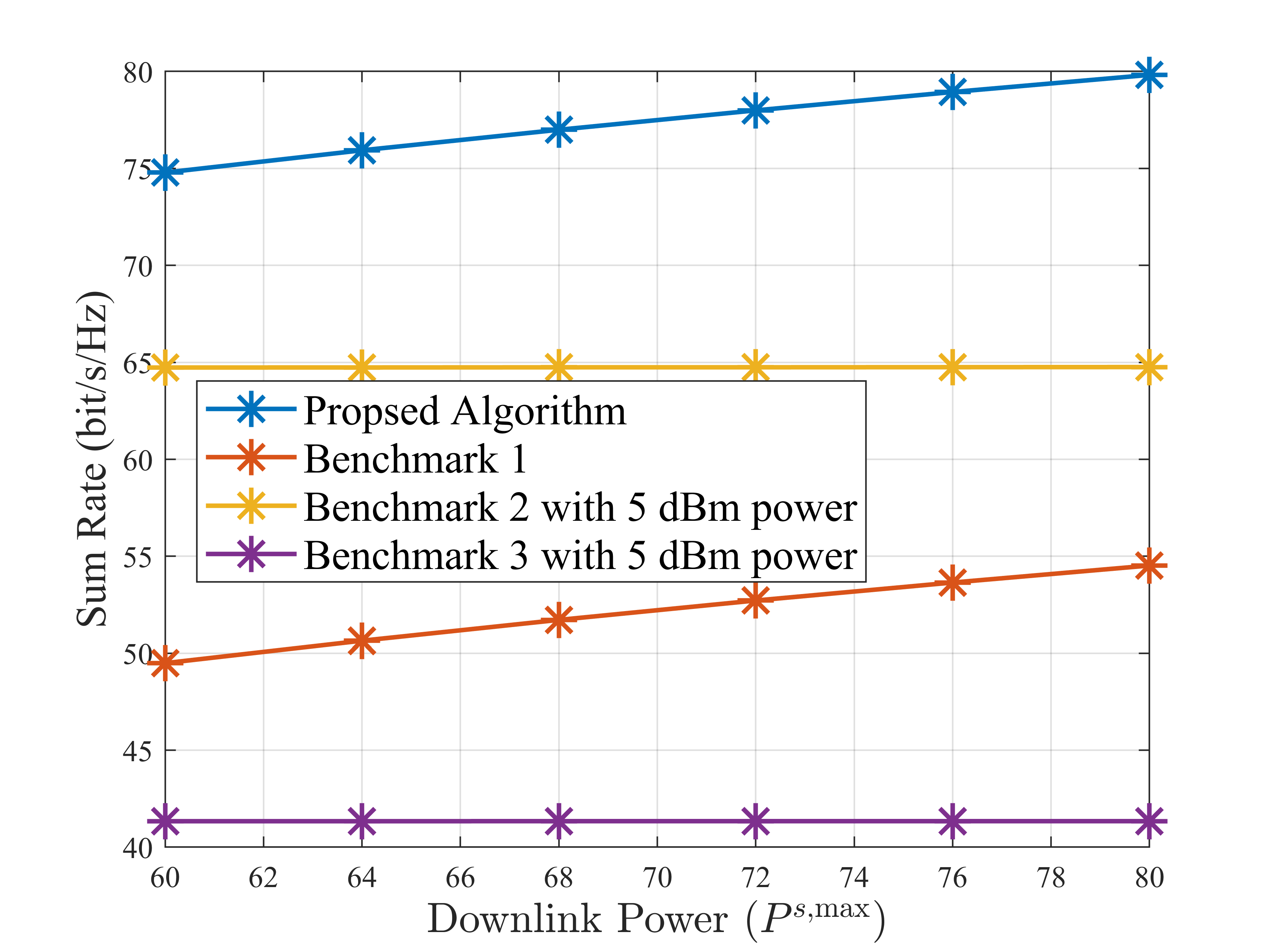}\hspace{10mm}
		%\caption{fig1}
\end{minipage}}
\caption{Sum rate under various downlink power \(P^{s,\max}\).}
\label{RatevsP}
\end{figure*}

Figure~\ref{RatevsP} depicts the sum rate versus the downlink transmit power by averaging over 100 random locations of devices. As the BS power increases, the RAQR-based scheme achieves a steadily rising sum rate, whereas the baseline of RF MIMO exhibits saturated performance. For benchmarks based on RF receiver, even though the harvested energy increases with higher downlink power, it remains negligible compared to the battery capacity of the equipment, leading to no benefits gained from the harvested energy. In contrast, for the BS equipped with RAQR, it is capable of capturing the weak signals from IoT devices, and thus the increasing harvested energy will significantly boost signal strength, resulting in the improved system performance. Furthermore, the proposed algorithm outperforms the benchmark schemes because it judiciously optimizes the uplink pilot power, uplink data power, downlink transmit power, and the power-splitting coefficients, which yields a more efficient allocation across energy harvesting and information transfer. 

\subsubsection{Effects of Required Rate}
\begin{figure*}[ht]
	\centering
	\subfigure[MRC (MRT) with \(P^{s,\max} = 60\) W and \(R^{{\rm req},D}_k = 1\), \(\forall k\).]{
		\begin{minipage}[t]{0.45\linewidth}
			\centering
			\includegraphics[width=2.75in]{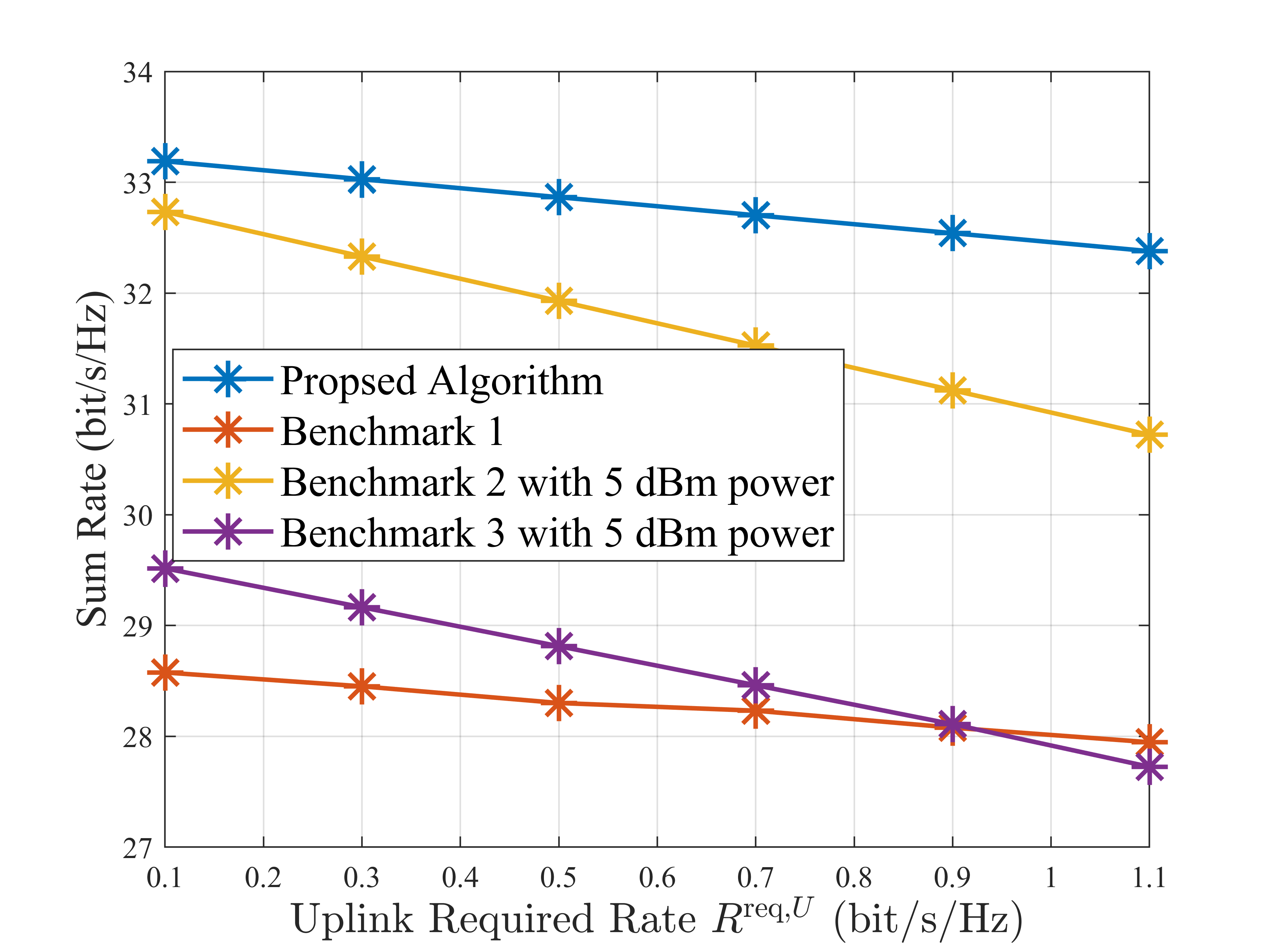}\hspace{10mm}
			%\caption{fig1}
	\end{minipage}}
	\quad%\includegraphics[width=2.75in]{cluster.png}}\hspace{10mm}
\subfigure[ZF with \(P^{s,\max} = 60\) W and \(R^{{\rm req},D}_k = 1\), \(\forall k\).]{
	\begin{minipage}[t]{0.45\linewidth}
		\centering
		\includegraphics[width=2.75in]{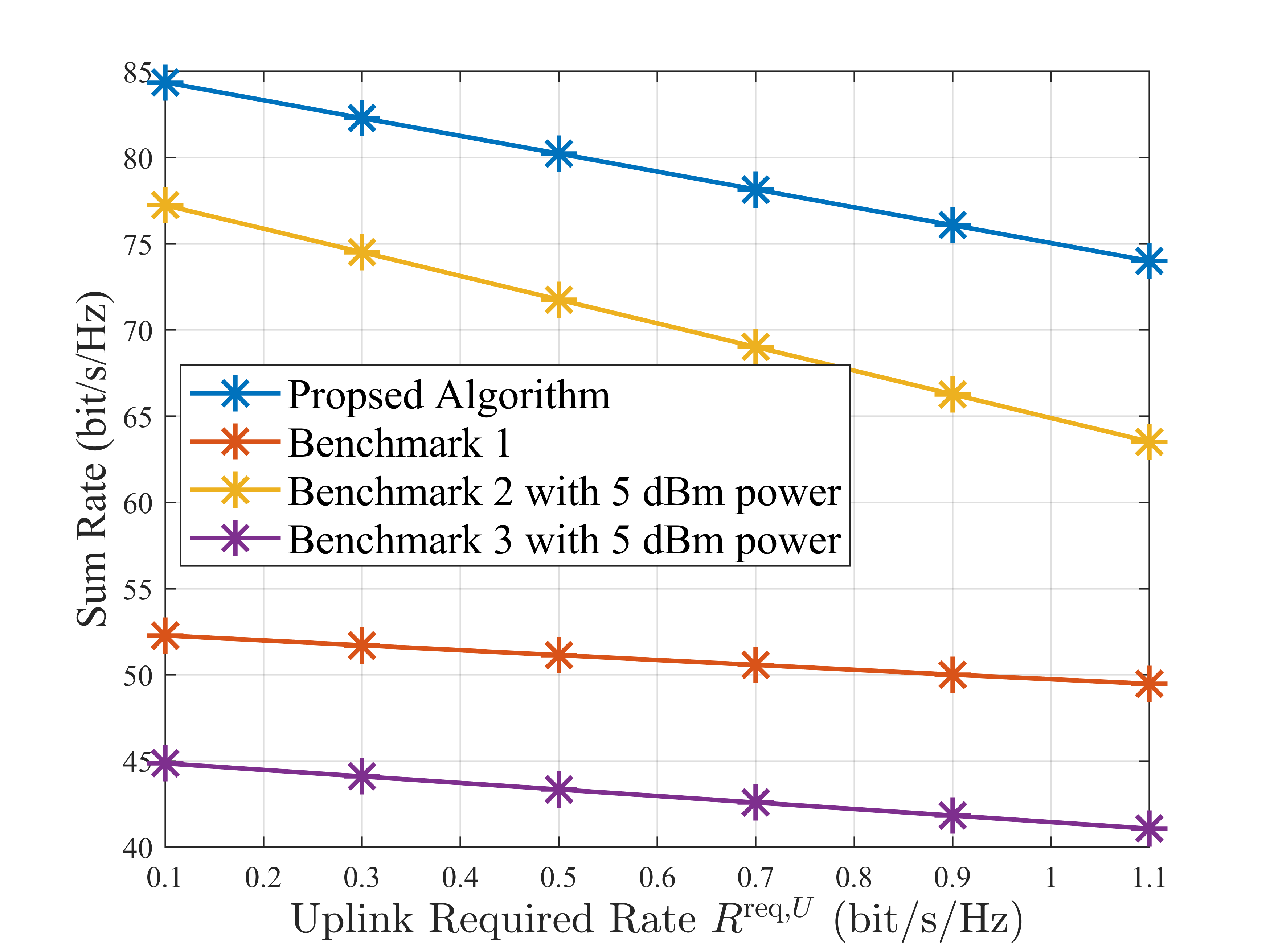}\hspace{10mm}
		%\caption{fig1}
\end{minipage}}
\caption{Sum rate under various downlink power \(P^{s,\max}\).}
\label{Ratevsreq}
\end{figure*}

As the downlink rate can be readily satisfied by optimizing the downlink power and power splitting coefficient, we focus on the impact of the required uplink rate on the system performance. Fig.~\ref{Ratevsreq} illustrates the sum rate under various uplink rates with \(P^{s,\max }= 60\) W. For the MRC/MRT case, the sum rate decreases as the required uplink rate increases, because more power must be allocated to devices with poor channels, which sacrifices resources from high-spectral-efficiency devices and lowers the system throughput. Regarding ZF,  the same trend appears in the baseline relying on RF receiver  but is much less pronounced in the RAQR-assisted system, since the requirement is more easily satisfied under ZF detection with RAQRs and the impact of the uplink rate constraint is reduced. Notably, the RAQR-based system maintains stable communication using only harvested energy, with no additional power consumption, whereas RF MIMO struggles to achieve comparable performance even with a 5 dBm uplink power budget. These results indicate that RAQR enables a more sustainable and performance-capable battery-free communication architecture for future IoT devices.

\subsubsection{Effects of Distance}
\begin{figure*}[ht]
	\centering
	\subfigure[MRC/MRT with \(P^{s,\max} = 60\)W, \(R^{{\rm req},U}_k = 1 \), \(R^{{\rm req},U}_k = 1\), \(\forall k\).]{
		\begin{minipage}[t]{0.45\linewidth}
			\centering
			\includegraphics[width=2.75in]{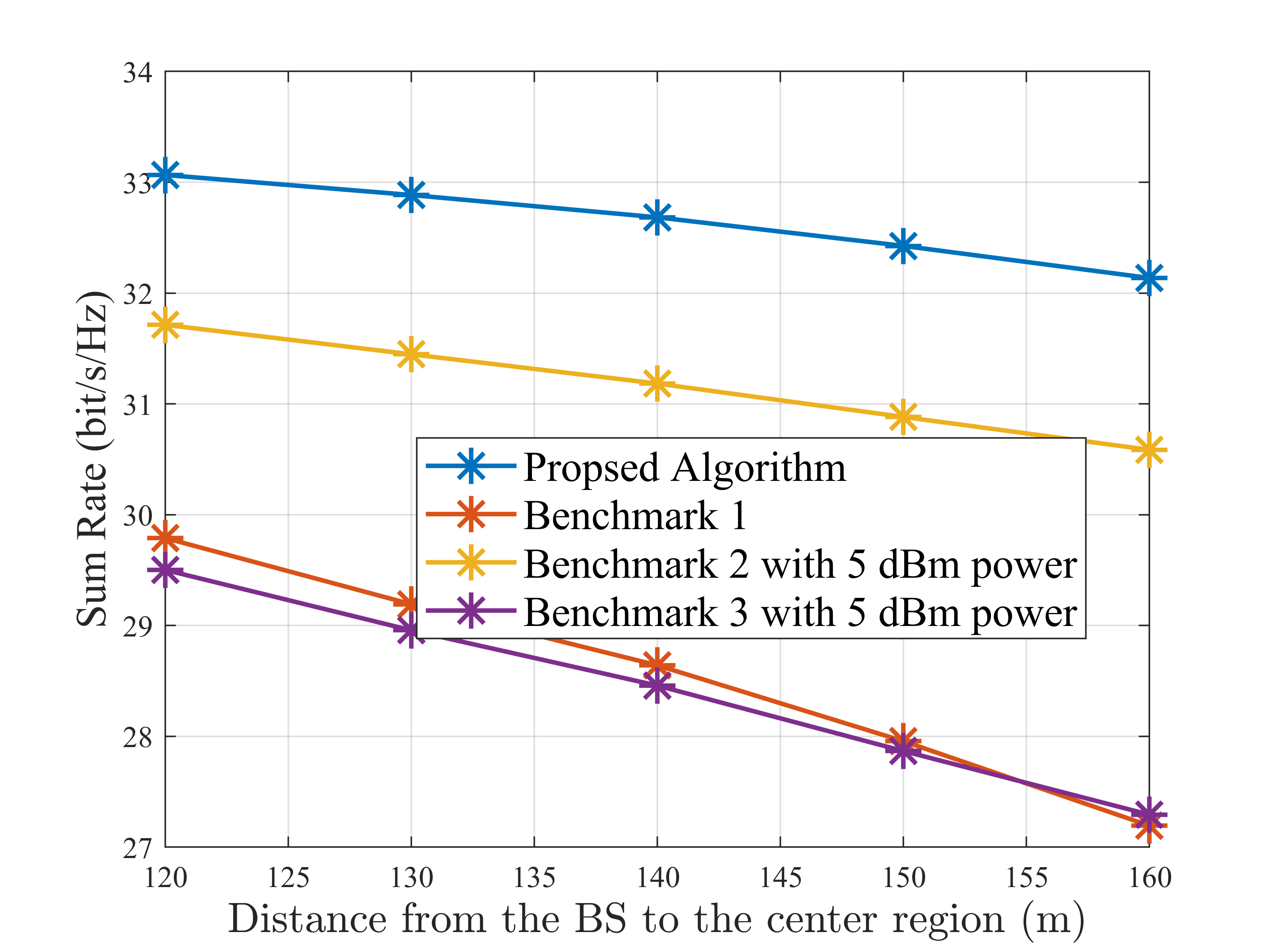}\hspace{10mm}
			%\caption{fig1}
	\end{minipage}}
	\quad%\includegraphics[width=2.75in]{cluster.png}}\hspace{10mm}
\subfigure[ZF with \(P^{s,\max} = 60\)W, \(R^{{\rm req},U}_k = 1 \), \(R^{{\rm req},U}_k = 1\), \(\forall k\).]{
	\begin{minipage}[t]{0.45\linewidth}
		\centering
		\includegraphics[width=2.75in]{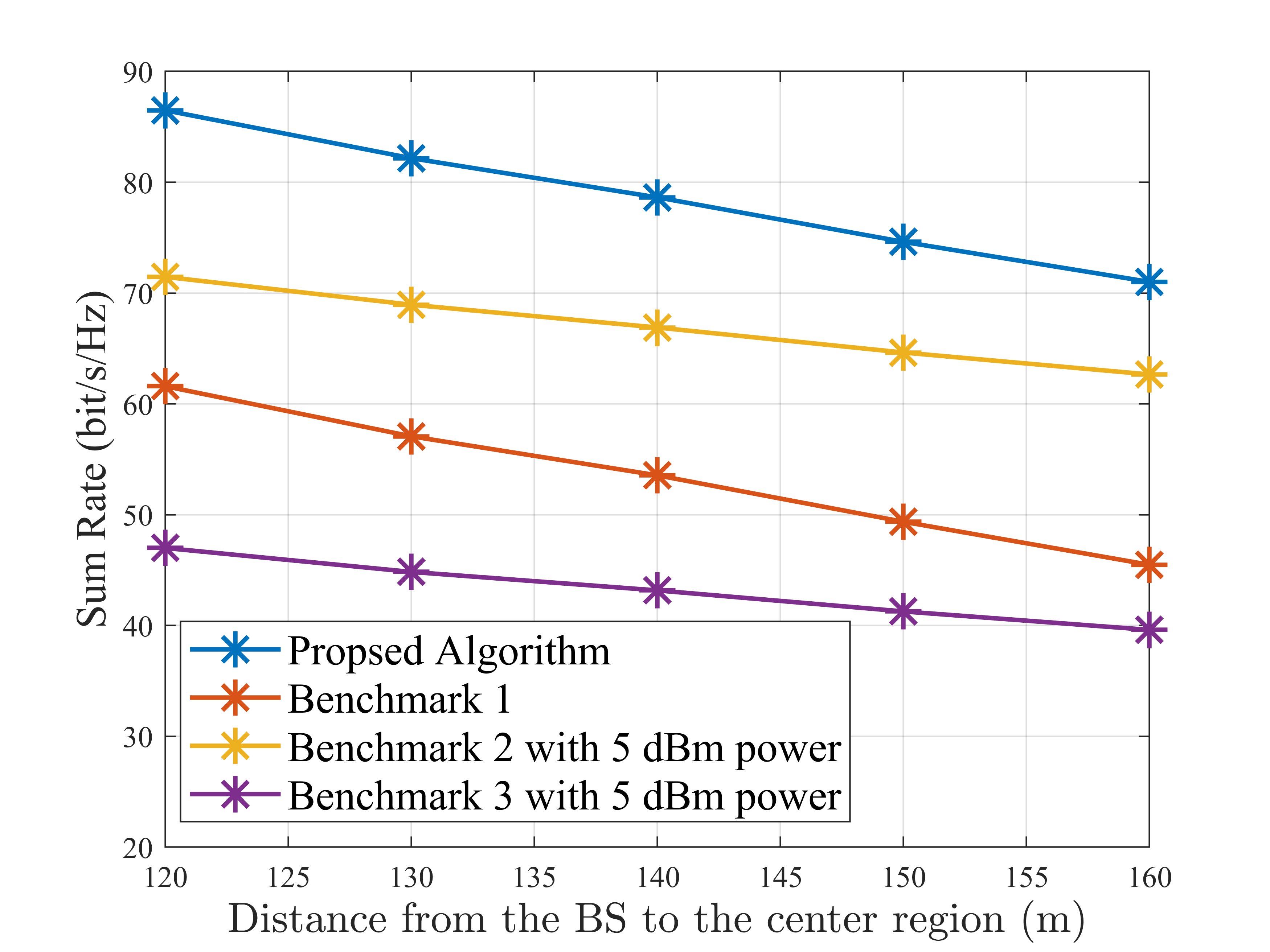}\hspace{10mm}
		%\caption{fig1}
\end{minipage}}
\caption{Sum rate under various propagation distance.}
\label{Ratevsdis}
\end{figure*}

Figure~\ref{Ratevsdis} examines the impact of propagation distance on system performance with a maximum downlink transmit power of \(P^{s,\max} = 60\) W. The sum rate decreases as the distance between the RAQR-equipped BS and the IoT devices increases, because the received signal power decays with distance, resulting in the degraded performance. We also observe that the performance gap between the proposed RAQR-assisted scheme and the baseline based on RF receiver gradually narrows with distance. This occurs because the harvested energy scales with the large-scale fading, which diminishes as the link distance grows, reducing the additional benefit obtained from energy harvesting. By contrast, Benchmarks~2 and~3 use battery-powered devices whose transmit power does not depend on the harvested energy. Nevertheless, the proposed algorithm and architecture outperform all benchmarks, demonstrating RAQR's potential for future powerless communication IoT devices.

\section{Conclusion}
In this paper, we investigated a SWIPT-enabled MIMO system with a RAQR, where the downlink simultaneously delivers information and energy, enabling battery-free IoT devices to transmit uplink pilots and data using harvested power. First, we derived closed-form lower bounds for the achievable rates and harvested energy under linear detection/precoding, providing a tractable expressions for power-allocation and channel-estimation tradeoffs. Building on these bounds, we formulated a joint uplink–downlink design that maximizes the sum rate subject to SWIPT constraints, resulting in a non-convex problem. To tackle this issue, the best monomial approximation and GP were performed iteratively and a rigorous convergence of proposed algorithm was provided. Finally, simulations confirmed the tightness of the bounds and demonstrated the superiority of the proposed algorithms over benchmark schemes, especially in low-power regimes. The results indicated that integrating Rydberg receivers with SWIPT MIMO can sustain device operation without extra batter and sacrificing spectral efficiency.

\begin{appendices}	
\section{Proof of theorem \ref{MRC_SINR_T}}
\label{Prooftheorem1}
By using \(\mathbf{c}_k = \Phi\mathbf{D}\mathbf{\hat h}_k\), we have
\begin{equation}
	\small 
	\mathbb{E}\{\sqrt{\rho p^d_k}\Phi\mathbf{c}^{H}_k\mathbf{D}\mathbf{h}_k\} = M\sqrt{\rho  p^d_k}|\Phi|^2 \frac{\rho \tau p^p_k\beta^2_k|\Phi|^2}{\rho\beta_k \tau p^p_k|\Phi|^2+\sigma^2}.
\end{equation}

Next, the power of the leaked signal can be expressed as
\begin{equation}
	\small 
	\begin{split}
		|\text{Ls}_k|^2 &= \mathbb{E}{\{|\sqrt{\rho p^d_k}\Phi\mathbf{c}^{H}_k\mathbf{D}\mathbf{h}_k|^2\}} - |\text{Ds}_k|^2 \\
		& = \rho p^d_k |\Phi|^4 \mathbb{E}\{\mathbf{\hat h}^H_k\mathbf{h}_k\mathbf{h}^H_k \mathbf{\hat h}_k\}-|\text{Ds}_k|^2 \\
		& = \rho p^d_k |\Phi|^4 (\frac{\rho\beta_k \tau p^p_k|\Phi|^2}{\rho\beta_k\tau p^p_k|\Phi|^2+\sigma^2})^2 \beta^2_k (M^2 + M) \\
		&+ \rho p^d_k |\Phi|^4 (\frac{\sqrt{\rho \tau p^p_k}\beta_k\Phi^H}{\rho\beta_k\tau p^p_k|\Phi|^2 + \sigma^2})^2 \sigma^2 \beta_kM - |\text{Ds}_k|^2 \\
		& = M \rho p^d_k |\Phi|^4\beta_k\frac{\rho \tau p^p_k \beta^2_k |\Phi|^2}{\rho\beta_k \tau p^p_k|\Phi|^2 + \sigma^2}. 
	\end{split}  
\end{equation}

The inter-device interference is given by
\begin{equation}
	\small 
	\begin{split}
		|\text{UI}_{k,k'}|^2 &= \mathbb{E}{\{|\sqrt{\rho p^d_{k'}}\Phi\mathbf{c}^{H}_k\mathbf{D}\mathbf{h}_{k'}|^2\}}\\
		& = \rho p^d_{k'} |\Phi|^4 \mathbb{E}\{\mathbf{\hat h}^H_{k}\mathbf{h}_{k'}\mathbf{h}^H_{k'} \mathbf{\hat h}_k\}\\
		& = M \rho p^d_{k'} |\Phi|^4 \beta_{k'} \frac{\rho \tau p^p_k \beta^2_k |\Phi|^2}{\rho\beta_k\tau p^p_k|\Phi|^2 + \sigma^2}.
	\end{split}  
\end{equation}

Similarly, the noise can be written as
\begin{equation}
	\small 
	\begin{split}
		|\text{N}_{k}|^2 &= \mathbb{E}{\{|\mathbf{c}^{H}_k\mathbf{w}|^2\}}\\
		& = M|\Phi|^2\sigma^2 \frac{\rho \tau p^p_k \beta^2_k |\Phi|^2}{\rho\beta_k\tau p^p_k|\Phi|^2 + \sigma^2}.
	\end{split}  
\end{equation}

Finally, we complete this proof by substituting the above results into (\ref{MRC_LB_rate}).

\section{Proof of theorem \ref{ZF_SINR_T}}
\label{Prooftheorem2}
By using \(\mathbf{C} =\Phi\mathbf{D}\mathbf{\hat H} [(\Phi\mathbf{D}\mathbf{\hat H})^H\Phi\mathbf{D}\mathbf{\hat H}]^{-1}\), we have
\begin{equation}
	\small 
	\mathbb{E}\{\sqrt{\rho p^d_k}\Phi\mathbf{c}^{H}_k\mathbf{D}\mathbf{h}_k\} = \sqrt{\rho p^d_k}.
\end{equation}

Then, the leaked signal based on ZF can be expressed as
\begin{equation}
	\small 
	\begin{split}
		|\text{Ls}_k|^2 &= \mathbb{E}{\{|\sqrt{\rho p^d_k}\Phi\mathbf{c}^{H}_k\mathbf{D}\mathbf{h}_k|^2\}} - |\text{Ds}_k|^2 \\
		& = \rho p^d_k  \mathbb{E}\{\mathbf{c}^H_k\mathbf{h}_k\mathbf{h}^H_k \mathbf{c}_k\}-|\text{Ds}_k|^2 \\
		& = \rho p^d_k \frac{\sigma^2}{(M-K) \rho \tau p^p_k \beta_k |\Phi|^2}. 
	\end{split}  
\end{equation}

Next, the inter-device interference and noise are given by
\begin{equation}
	\small 
	\begin{split}
		|\text{UI}_{k,k'}|^2 &= \mathbb{E}{\{|\sqrt{\rho p^d_{k'}}\Phi\mathbf{c}^{H}_k\mathbf{D}\mathbf{h}_{k'}|^2\}}\\
		& = \rho p^d_{k'} \mathbb{E}\{\mathbf{c}^H_{k}\mathbf{h}_{k'}\mathbf{h}^H_{k'} \mathbf{c}_k\}\\
		& = \rho p^d_{k'}  \frac{ \frac{\beta_{k'}\sigma^2}{\rho\beta_{k'}\tau p^p_{k'}|\Phi|^2 + \sigma^2}}{(M-K)\frac{\rho \tau p^p_k \beta^2_k |\Phi|^2}{\rho\beta_k\tau p^p_k|\Phi|^2 + \sigma^2}},
	\end{split}  
\end{equation}
and
\begin{equation}
	\small 
	\begin{split}
		|\text{N}_{k}|^2 &= \mathbb{E}{\{|\mathbf{c}^{H}_k\mathbf{w}|^2\}}\\
		& = \frac{\sigma^2 }{(M-K)|\Phi|^2 \frac{\rho \tau p^p_k \beta^2_k |\Phi|^2}{\rho\beta_k\tau p^p_k|\Phi|^2 + \sigma^2}}.
	\end{split}  
\end{equation}

By combining the results, the proof of Theorem \ref{ZF_SINR_T} is completed.

\section{Proof of Theorem~\ref{MRCapprox}}\label{ProofMRCappro}

Define the $2K$ positive terms
\begin{equation}
	u^{(1)}_{k'}(\mathbf p^s,p^p_k) \triangleq B_k\,p^s_{k'}, 
	u^{(2)}_{k'}(\mathbf p^s,p^p_k) \triangleq p^p_k\,A_{k'}\,p^s_{k'}, \forall k',
\end{equation}
and denote their sum as
\begin{equation}
	S(\mathbf p^s,p^p_k)\triangleq B_k\sum_{k'=1}^K p^s_{k'}+p^p_k\sum_{k'=1}^K A_{k'}p^s_{k'}.
\end{equation}
With the given fixed point $(\hat{\mathbf p}^s,\hat p^p_k)$, we have
\begin{equation}
\alpha^{(1)}_{k'} \triangleq \frac{u^{(1)}_{k'}(\hat{\mathbf p}^s,\hat p^p_k)}{S(\hat{\mathbf p}^s,\hat p^p_k)},
\qquad
\alpha^{(2)}_{k'} \triangleq \frac{u^{(2)}_{k'}(\hat{\mathbf p}^s,\hat p^p_k)}{S(\hat{\mathbf p}^s,\hat p^p_k)},
\end{equation}
where \(S(\hat{\mathbf p}^s,\hat p^p_k)\), \(u^{(1)}_{k'}(\hat{\mathbf p}^s,\hat p^p_k)\), and \(u^{(2)}_{k'}(\hat{\mathbf p}^s,\hat p^p_k)\) are obtained by setting \(p^p_k = \hat p^p_k\) and \({\mathbf p}^s = \hat{\mathbf p}^s\).
By using the inequality of arithmetic and geometric means, we have
\begin{equation}
\begin{split}
&S(\mathbf p^s,p^p_k)\\
	\ge&
	S(\hat{\mathbf p}^s,\hat p^p_k)\;
	\prod_{k'=1}^K\!\Bigg(\frac{u^{(1)}_{k'}(\mathbf p^s,p^p_k)}{u^{(1)}_{k'}(\hat{\mathbf p}^s,\hat p^p_k)}\Bigg)^{\alpha^{(1)}_{k'}}  \prod_{k'=1}^K\!\Bigg(\frac{u^{(2)}_{k'}(\mathbf p^s,p^p_k)}{u^{(2)}_{k'}(\hat{\mathbf p}^s,\hat p^p_k)}\Bigg)^{\alpha^{(2)}_{k'}} \\
	=&
	\underbrace{\frac{S(\hat{\mathbf p}^s,\hat p^p_k)}{
			\prod_{k'}\big(u^{(1)}_{k'}(\hat{\mathbf p}^s,\hat p^p_k)\big)^{\alpha^{(1)}_{k'}}
			\prod_{k'}\big(u^{(2)}_{k'}(\hat{\mathbf p}^s,\hat p^p_k)\big)^{\alpha^{(2)}_{k'}}}}_{\delta_k} 
\\
	\times&
	\prod_{k'=1}^K \big(u^{(1)}_{k'}(\mathbf p^s,p^p_k)\big)^{\alpha^{(1)}_{k'}}
	\prod_{k'=1}^K \big(u^{(2)}_{k'}(\mathbf p^s,p^p_k)\big)^{\alpha^{(2)}_{k'}}.
\end{split}
\end{equation}
Therefore, we complete this proof by defining
\begin{equation}
	\begin{split}
		\omega_k^p
		&=\sum_{k'=1}^K \alpha^{(2)}_{k'}
		=\frac{\hat p_k^p\sum\limits_{k'=1}^K A_{k'}\hat p^s_{k'}}
		{S(\hat{\mathbf p}^s,\hat p^p_k)},
		\\
		\omega^s_{k,k'}
		&=\alpha^{(1)}_{k'}+\alpha^{(2)}_{k'}
		=\frac{B_k\hat p^s_{k'}+\hat p_k^pA_{k'}\hat p^s_{k'}}
		{S(\hat{\mathbf p}^s,\hat p^p_k)}.
	\end{split}
\end{equation}

\end{appendices}	

%\subsection{Effect of Blocklength}
%We investigate the effect of blocklength on the weighted sum rate, as illustrated in Fig. . 
%\bibColoredItems{black}{zhi2022power,wu2020joint,guo2022uplink,loyka2001channel}

\bibliographystyle{IEEEtran}
\bibliography{myref}

\end{document}